\newcommand{\ieee}{} %
\ifdefined\ieee
\documentclass[conference,compsoc]{support/IEEEtran}
\IEEEoverridecommandlockouts
\else
\documentclass[letterpaper,twocolumn,10pt]{article}
\usepackage{support/usenix_v3}
\fi

\usepackage{xspace}
\usepackage{textcomp}

\usepackage{enumitem}
\usepackage[normalem]{ulem}

\usepackage{etoolbox}

\usepackage[nomessages]{fp}

\usepackage{amsmath,amssymb,amsfonts}
\usepackage{dsfont}
\usepackage{xcolor}

\usepackage{layouts}
\usepackage{caption}
\usepackage{subcaption}
\usepackage{graphicx}
\usepackage{float}

\usepackage{booktabs}
\usepackage{tabularx}
\usepackage{array}
\usepackage{arydshln}

\usepackage{cite}
\ifdefined\ieee
\usepackage[]{hyperref}
\hypersetup{
  colorlinks,
  linkcolor={green!80!black},
  citecolor={red!70!black},
  urlcolor={blue!70!black}
}
\fi
\usepackage[capitalize]{cleveref}

\usepackage{algorithmic}
\usepackage[
    lambda,
    operators,
    advantage,
    sets,
    adversary,
    landau,
    probability,
    notions,
    logic,
    ff,
    mm,
    primitives,
    events,
    oracles,
    complexity,
    asymptotics,
    keys]{support/cryptocode}

\def\BibTeX{{\rm B\kern-.05em{\sc i\kern-.025em b}\kern-.08em
    T\kern-.1667em\lower.7ex\hbox{E}\kern-.125emX}}
\newtoggle{comments} %
\newtoggle{people} %
\newtoggle{newpage} %
\newtoggle{camera} %

\togglefalse{newpage}

\iftoggle{newpage}{ 
\newcommand{\newpagetoggle}{\newpage}{}
}{
\newcommand{\newpagetoggle}{}{}
}

\setlist[itemize]{noitemsep,nosep,leftmargin=*}
\setlist[enumerate]{noitemsep,nosep,leftmargin=*}

\iftoggle{comments}{
\newcommand{\lhm}[1]{\textcolor{red}{LHM:#1}}
\renewcommand{\aa}[1]{\textcolor{orange}{AA:#1}}
\newcommand{\zhq}[1]{\textcolor{orange}{ZHQ:#1}}
\renewcommand{\sc}[1]{\textcolor{orange}{SC:#1}}
\newcommand{\veg}[1]{\textcolor{purple}{VEG:#1}}
\newcommand{\baf}[1]{\textcolor{purple}{BAF:#1}}
}{
\newcommand{\lhm}[1]{\ignorespaces}
\newcommand{\baf}[1]{\ignorespaces}
\renewcommand{\aa}[1]{\ignorespaces}
\newcommand{\zhq}[1]{\ignorespaces}
\renewcommand{\sc}[1]{\ignorespaces}
\newcommand{\veg}[1]{\ignorespaces}
}

\newcommand{\soutthick}[1]{
\renewcommand{\ULthickness}{1.4pt}\sout{#1}
\renewcommand{\ULthickness}{.4pt}}

\definecolor{blue-violet}{rgb}{0.54, 0.17, 0.89}
\definecolor{forest-green}{rgb}{0.13, 0.54, 0.13}
\definecolor{soft-blue}{rgb}{0.29, 0.61, 0.82}

\iftoggle{comments}{
\long\def\zhq#1{ {\bf ZHQ: } [{\color{red} \em #1}]}
\newcommand{\todo}[1]{\textcolor{red}{TODO:#1}\xspace}
\newcommand{\todoSilent}[1]{\textcolor{red}{#1}\xspace}
\newcommand{\com}[1]{\textcolor{orange}{#1}}
\newcommand{\comcom}[1]{}

}{
\long\def\zhq#1{}
\newcommand{\todoSilent}[1]{}
\newcommand{\todo}[1]{}
\newcommand{\com}[1]{}
\newcommand{\comcom}[1]{}

}

\iftoggle{camera}{

\newcommand{\remove}[1]{\textcolor{soft-blue}{\soutthick{#1}\xspace}}
\newcommand{\removefootnote}[1]{\footnote{\textcolor{soft-blue}{\soutthick{#1}\xspace}}}
}{

\newcommand{\remove}[1]{}
\newcommand{\removefootnote}[1]{}
}

\iftoggle{comments}{
\newcommand{\secnotes}[1]{
\textcolor{purple}{
--------------------------- Section Notes ---------------------------}
#1
\noindent \textcolor{purple}{\hrule}\vspace{0.2cm}}}
{
\newcommand{\secnotes}[1]{}
}

\newcommand{\eg}{e.g.,\xspace}
\newcommand{\ie}{i.e.,\xspace}

\crefformat{section}{\S#2#1#3}
\Crefformat{section}{\S#2#1#3}
\newcommand{\parhead}[1]{\noindent \textbf{#1}}

\newcommand{\materialslink}{https://github.com/dedis/trip-usability}

\createprocedureblock{gameblock}{center,boxed}{}{}{linenumbering}
\renewcommand{\pcadvantagesuperstyle}[1]{\mathrm{#1}}

\let\OLDthebibliography\thebibliography
\renewcommand\thebibliography[1]{
  \OLDthebibliography{#1}
  \setlength{\parskip}{0pt}
  \setlength{\itemsep}{0pt plus 0.3ex}
}

\begin{document}
\bstctlcite{support/MyBSTcontrol}

\title{E-Vote Your Conscience: \\
Perceptions of Coercion and Vote Buying, \\and
the Usability of Fake Credentials in Online Voting}
\author{
\IEEEauthorblockN{
Louis-Henri Merino\IEEEauthorrefmark{1},
Alaleh Azhir\IEEEauthorrefmark{2},
Haoqian Zhang\IEEEauthorrefmark{1},
Simone Colombo\IEEEauthorrefmark{1}, \\
Bernhard Tellenbach\IEEEauthorrefmark{3},
Vero Estrada-Galiñanes\IEEEauthorrefmark{1},
Bryan Ford\IEEEauthorrefmark{1}}
\IEEEauthorblockA{\IEEEauthorrefmark{1}EPFL \quad \IEEEauthorrefmark{2}MIT \quad \IEEEauthorrefmark{3}Armasuisse}
} %

\IEEEaftertitletext{\vspace{-1\baselineskip}\centering\small \emph{This paper is the extended version of a work published in\\ the proceedings of the 45th IEEE Symposium on Security and Privacy, May 2024.\vspace{1\baselineskip}}}

\maketitle

\thispagestyle{plain}
\pagestyle{plain}

\begin{abstract}
Online voting is attractive for convenience and accessibility,
but is more susceptible to 
voter coercion and vote buying than 
in-person voting.
One mitigation is to give voters \emph{fake voting credentials}
that they can yield to a coercer.
Fake credentials appear identical to real ones,
but cast votes that are silently omitted from the final tally.
An important unanswered question
is how ordinary voters perceive such a mitigation:
whether they could understand and use fake credentials,
and whether the coercion risks justify the costs of mitigation.
We present the first systematic study of these questions,
involving 150 diverse individuals in Boston, Massachusetts.
All participants ``registered'' and ``voted'' in a mock election:
120 were exposed to coercion resistance via fake credentials,
the rest forming a control group.
Of the 120 participants exposed to fake credentials,
96\% understood their use.
53\% reported that they would create fake credentials
in a real-world voting scenario,
given the opportunity.
10\% mistakenly voted with a fake credential, however.
22\% 
reported either personal experience with or direct knowledge
of coercion or vote-buying incidents.
These latter participants rated the coercion-resistant system
essentially as trustworthy as in-person voting via
hand-marked paper ballots.
Of the 150 total participants to use the system,
87\% successfully created their credentials without assistance;
83\% both successfully created and properly used their credentials.
Participants give a System Usability Scale score of 70.4,
which is slightly above the industry's average score of 68.
Our findings appear to support
the importance of the coercion problem in general,
and the promise of fake credentials as a possible mitigation,
but user error rates remain an important usability challenge
for future work.

\end{abstract}

\section{Introduction}\label{sec:intro}
Remote electronic (online) voting systems
promise convenience and increased voter turnout~\cite{
ehin2022InternetVotingEstonia,
besselaar2003EVotingExperimentMultiCountry}.
Online voting is particularly useful
to overseas voters~\cite{
jefferson2004SERVEOverseasVoters}
or in crises such as a pandemic~\cite{
krimmer2021InternetVotingPandemic}.
One important development in electronic voting
is universal verifiability,
which allows anyone (not just election officials and observers)
to verify that votes have been tallied correctly,
while protecting voter
privacy~\cite{benaloh1987VerifiableSecretBallot,adida2008Helios,cortier2016SoKVerifiabilityEVoting,kusters2020Ordinos,kho2022ReviewEVotingSystems}.

\emph{Individual verifiability}
measures~\cite{benaloh2007BenalohChallenge,
cortier2023BeleniosCaI}
attractively enable voters to verify that their votes
are cast as intended.
Individually-verifiable \emph{receipts}
unfortunately make voting more susceptible
to voter coercion~\cite{
juels2010CoercionResistantElections,
benaloh1994ReceiptfreeSecretBallot}.\footnote{
We use the term ``coercion'' broadly to indicate any form of
undue influence, including vote buying
and voter intimidation.}
An abusive partner or other coercer
might demand the voter's receipts,
for example~\cite{
park2021BadToWorse}.
These receipts could also enable unscrupulous well-funded
actors to buy votes at scale through
anonymously-funded smart contracts~\cite{
daian2018DarkDAOs}.
To resist such attacks,
a secure election system must prevent coercive adversaries
from knowing whether a voter complied with their demands,
even if the voter is willing to comply,
\eg in return for financial compensation~\cite{
juels2010CoercionResistantElections,
kusters2010CoercionResistanceDefinition}.

Most online voting systems
lack coercion resistance~\cite{2021SwissPostProofs,
bell2013STARVote,
adida2008Helios}.
\emph{Deniable re-voting} permits a voter
to override a coerced vote with a new vote
cast later~\cite{achenbach2015JCJDeniableRevoting,krips2019CRSummary,
lueks2020VoteAgain},
but
is vulnerable to coercers who
can supervise voters or hold onto their credentials or voting devices
until the election closes~\cite{park2021BadToWorse}.
Estonia's online voting system employs
deniable re-voting~\cite{kulyk2020HumanFactorsCoercion,
springall2014EstoniaEVotingAnalysis,achenbach2015JCJDeniableRevoting},
but lacks universal verifiability.

A strategy proposed
by Juels, Catalano and Jakobsson (JCJ)~\cite{
juels2010CoercionResistantElections}
enables voters to create,
alongside their real voting credential,
fake credentials which cast votes that
do not count.
Fake credentials present usability concerns,
however,
such as whether voters can distinguish their real credential
from fake ones,
or can create a fake
credential while under coercion~\cite{
kulyk2020HumanFactorsCoercion,neumann2012CivitasRealWorld}.
While prior work has discussed the usability of
fake credentials~\cite{kulyk2020HumanFactorsCoercion,
estaji2020UsableCRSmartCards,
neumann2012CivitasRealWorld
},
only Neto et al.~\cite{
neto2018CredentialsDistributionUsability}
performed a user study on this topic.
Their study
involved only university-affiliated participants, however.
Further, the voting process they studied
lacked individual verifiability:
voters could not check
whether the purportedly ``real'' credential they were issued
was in fact real (as opposed to fake).
Prior work also leaves other unanswered questions,
such as whether ordinary voters
even comprehend the coercion threat
or believe it is important.

To fill this gap,
we conducted a study with 150 individuals,
recruited at a suburban park in Boston, Massachusetts,
to examine whether voters might plausibly find
coercion-resistant online voting with fake credentials
to be usable and trustworthy.
120 of these participants underwent a credentialing
process to obtain real and
fake credentials.\footnote{
The study used the term ``test credentials''
instead of ``fake credentials''
to avoid negative connotations that
fake credentials are invalid or inherently bad.
The registration process suggests to users that ``test credentials''
may also be used for purposes other than coercion resistance,
such as to test the voting system,
or to share with friends or family for educational purposes.}
Participants then cast a mock vote using, at least,
their real credential.
The remaining 30 participants engage with the same
system, but without any exposure to fake credentials.
Participants conclude the study by
completing a survey asking them to share their
experiences and views of the system,
as well as their perspectives on and any experiences with coercion in general.
Our institutional review board approved the study;
Section~\ref{sec:methods} discusses ethics considerations.

The credentialing process is an interactive user-interface
prototype of Trust-limiting In-Person Registration (TRIP)~\cite{TRIP},
a voter-verifiable registration system for coercion-resistant
online voting via fake credentials.
Unlike prior coercion-resistant systems
either deployed~\cite{ehin2022InternetVotingEstonia,
	besselaar2003EVotingExperimentMultiCountry}
or subject to user studies~\cite{
	neto2018CredentialsDistributionUsability},
TRIP ensures that a compromised registrar
cannot undetectably manipulate elections by
secretly keeping real credentials for themselves,
leaving voters with only fake credentials.
To achieve verifiability without producing receipts usable for coercion,
TRIP's registration kiosk produces interactive zero-knowledge proof transcripts,
all of which are valid and checkable,
but are \emph{true} proofs only in real credentials,
and are false proofs in fake credentials.
Study participants used TRIP
to create real and fake credentials,
then used an Android device we supplied
to cast a mock non-political vote.
A random subset of participants were silently exposed to
a ``compromised'' kiosk that issued only fake credentials.
Because individual verifiability depends on voters being able to detect
a compromised kiosk -- not just in theory but \emph{in practice} --
our study sheds light for the first time on whether ordinary voters
can effectively obtain verifiability \emph{and} coercion resistance at once.

Using the data we collected from the study,
we address the following four central questions:
\begin{enumerate}[noitemsep]
\item What are voters'
perceptions of and experiences
with the coercion threat in general?
\item How likely will voters trust a coercion-resistant online voting system,
versus other voting methods?
\item Can voters use a voter-verifiable
credentialing process to create their real and fake credentials,
and identify deviations to ensure voter verifiability?
\item Can voters understand and use fake credentials,
casting their intended votes with their
real one?
\end{enumerate}

This is not a longitudinal study,
so it cannot assess issues such as
whether voters can recall the distinction
between their real and fake credentials
over an extended time period.
This is one of several limitations
we detail in \cref{sec:limitations}.

\noindent
This paper makes the following key contributions:
\begin{itemize}
    \item The first study to systematically investigate whether
        voters find coercion-resistant online voting with fake
        credentials to be understandable, usable and trustworthy.
    \item The first study that assesses whether voters can use
        a credentialing process that requires voters to identify and
        report a misbehaving kiosk to ensure voter verifiability.
\end{itemize}
Our prototype, which includes a user interface mockup of TRIP
	and an Android application that simulates activation and voting,
	is available at
        \href{https://github.com/dedis/trip-usability}{github.com/dedis/trip-usability}.

\section{Background}\label{sec:background}
This section introduces coercion-resistant voting systems,
TRIP, and the metrics we used in our study.

\subsection{Online Voting Systems}
Nearly all online voting systems~\cite{adida2008Helios,
2021SwissPostProofs,benaloh1987VerifiableSecretBallot}
strive for a minimum of verifiability and voter privacy:
establishing vote confidentiality,
while offering (publicly) verifiable election results.
Substantial research seeks to achieve other
desirable properties as well, such as cast-as-intended~\cite{
cortier2023BeleniosCaI,
benaloh2007BenalohChallenge},
where voters are convinced that their encrypted
ballot contains their intended vote.
JCJ~\cite{juels2010CoercionResistantElections}
proposed and formally defined coercion resistance
as another desirable system property:
in brief, the inability of an adversary
to confirm whether a coerced voter has complied with their demands,
\emph{even if the voter wishes to do so}.
While JCJ suggested the use of fake voting credentials,
other works~\cite{
achenbach2015JCJDeniableRevoting,
chaum2020VoteXXRemoteVoting,backes2013UsingMobileDevice}
have since proposed other strategies due
to usability concerns with fake credentials~\cite{
estaji2020UsableCRSmartCards}.
This paper focuses on these concerns.

A typical online voting system interacts with
voters, an election authority, and observers,
and involves setup, registration, voting, and tallying phases.
Performance-oriented systems research tends to focus on tallying:
reducing the costs of shuffling and counting ballots,
including the removal of fake ballots in
coercion-resistant voting systems.
For usability, however,
registration and voting
are the more crucial stages
because they directly involve voters.
This study focuses primarily on
registration, where voters engage with
the election authority to generate their
voting materials.

\subsection{Trust-limiting In-Person Registration}
\label{sec:background:TRIP}
\begin{figure*}[t]
    \centering
    \begin{subfigure}[b]{\textwidth}
        \includegraphics[width=\linewidth]{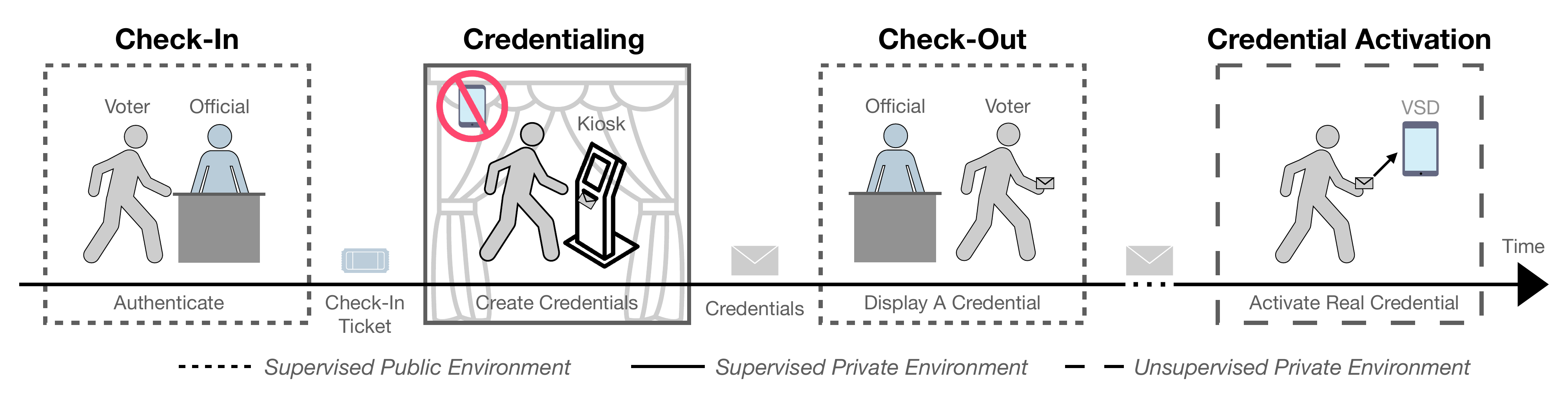}
        \caption{\textbf{TRIP Voter Workflow.}
            The voter 
          (1) checks in with an official by authenticating
          themselves to receive a check-in ticket,
          (2) enters a supervised private environment
            to create their credentials using a kiosk that is
            unlocked by their check-in ticket,
          (3) checks out with an official by displaying
            the public part of any one of their paper credentials 
            (Check-Out Ticket -- \cref{fig:TRIP:credentials:transport}), and
          (4) activates their credentials on their voter supporting device (VSD).
        }
        \label{fig:TRIP:voter-workflow}
    \end{subfigure}
\end{figure*}
\begin{figure}[t]
    \begin{subfigure}[b]{0.24\linewidth}
        \includegraphics[width=\linewidth]{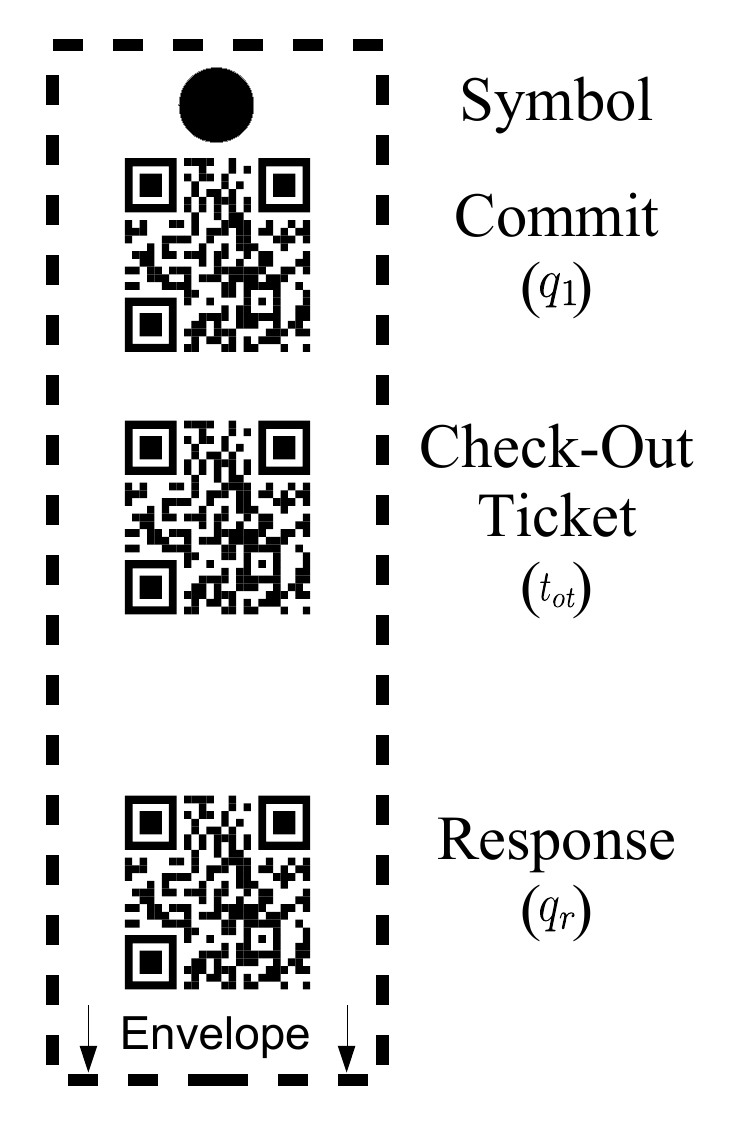}
        \caption{\textbf{Receipt}}
        \label{fig:TRIP:credentials:receipt}
    \end{subfigure}
    \begin{subfigure}[b]{0.24\linewidth}
        \includegraphics[width=\linewidth]{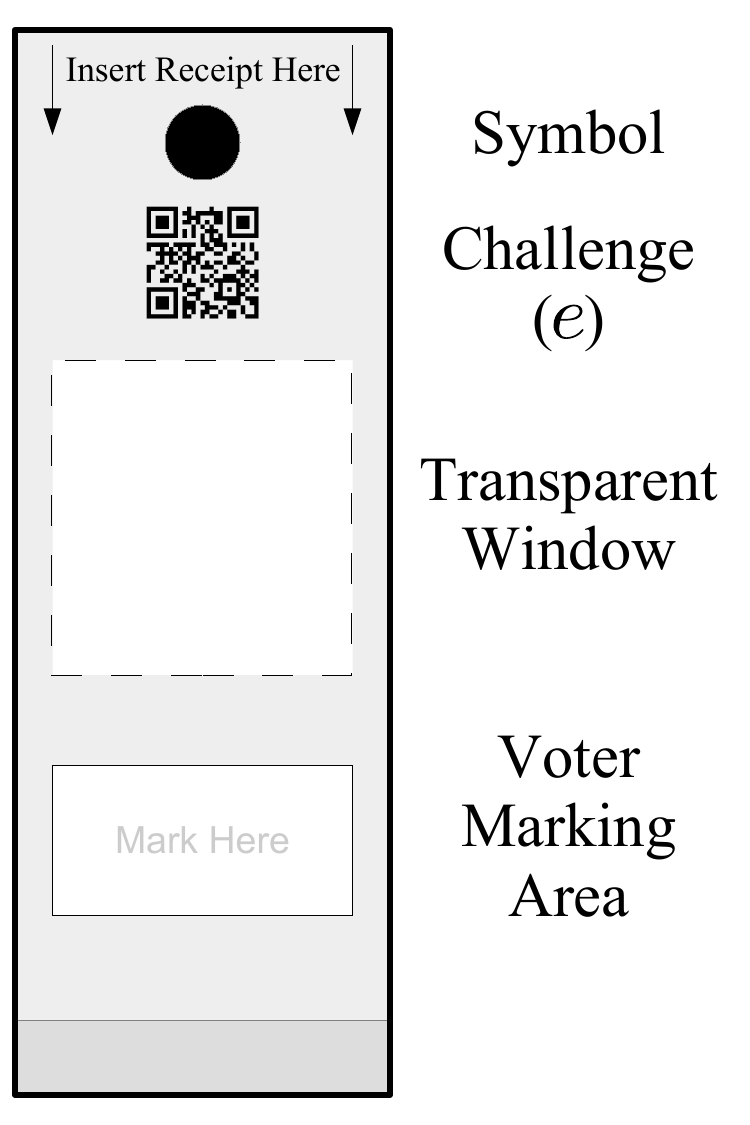}
        \caption{\textbf{Envelope}}
        \label{fig:TRIP:credentials:envelope}
    \end{subfigure}
    \begin{subfigure}[b]{0.24\linewidth}
        \includegraphics[width=\linewidth]{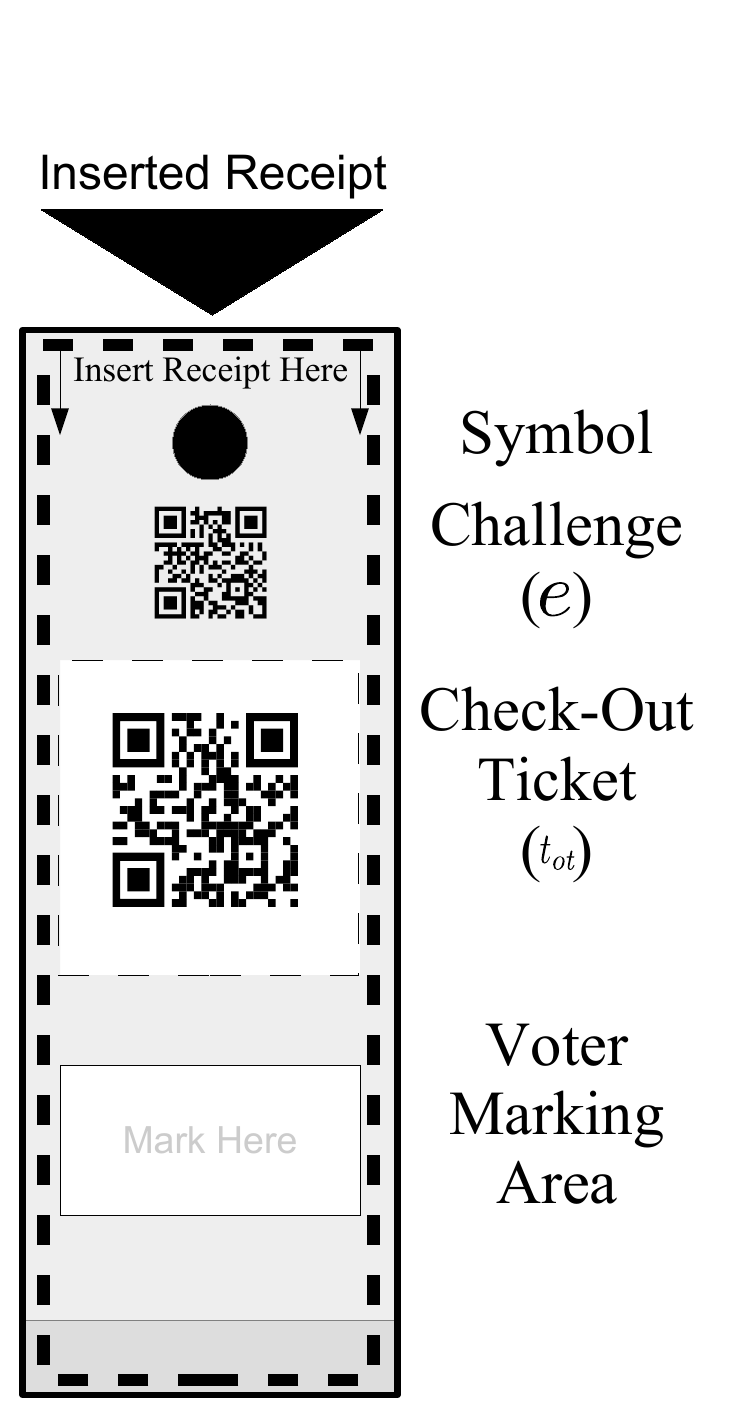}
        \caption{\centering \textbf{Transport}}
        \label{fig:TRIP:credentials:transport}
    \end{subfigure}
    \begin{subfigure}[b]{0.24\linewidth}
        \includegraphics[width=\linewidth]{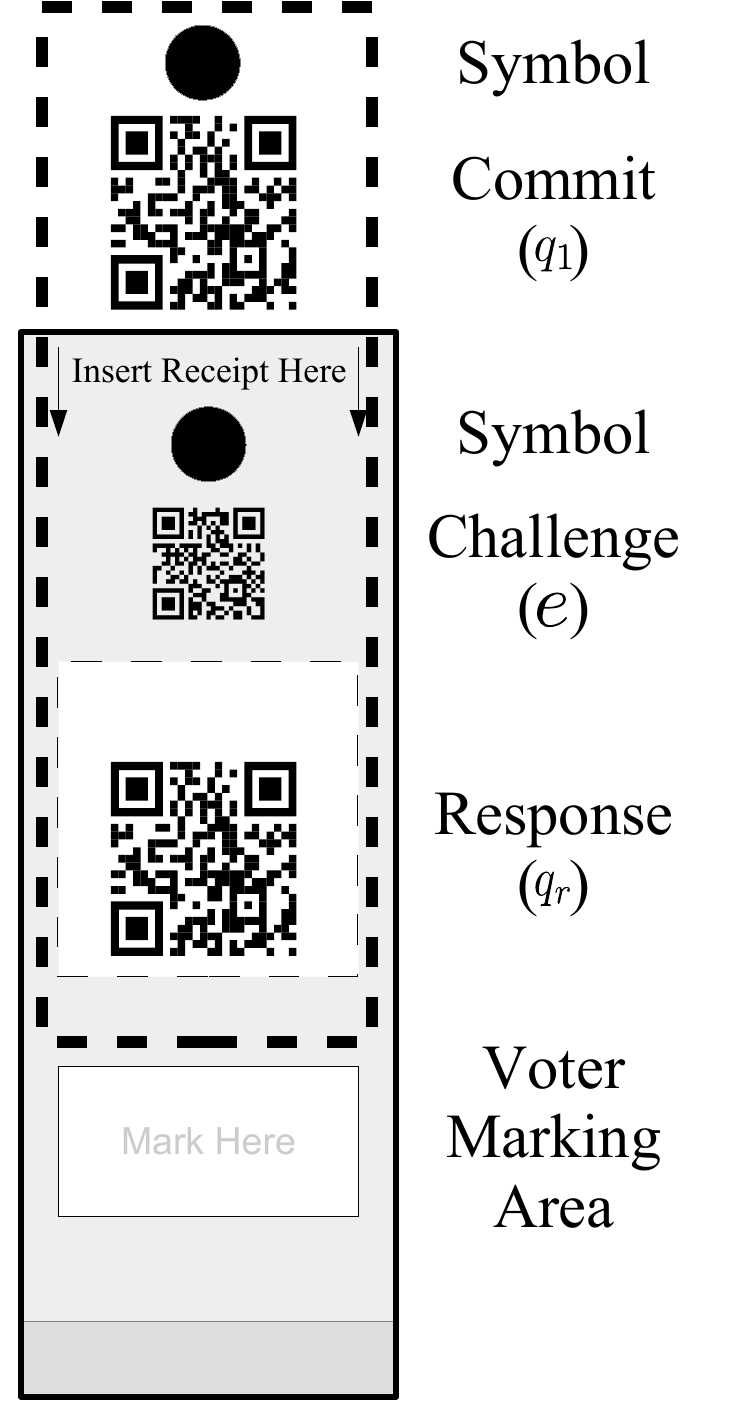}
        \caption{\centering \textbf{Activate}}
        \label{fig:TRIP:credentials:activate}
    \end{subfigure}
    \caption{\textbf{TRIP Paper Credential}. Figures (a) and (b) present 
    the paper credential's elements and Figures (c) and (d) present the 
	paper credential's transport and activate states.}
\end{figure}

Since the introduction of coercion resistance,
most works have focused on the tallying process~\cite{Krivoruchko2007CRVoterRegistration,
spycher2012JCJLinearTime,araujo2010PracticalSecureCoercionResistant,
benaloh2009CRSingleTV,weber2007CRLinear},
leaving a key challenge unresolved:
the development of a usable voter registration system
that issues \emph{voter-verifiable}
real and fake credentials.
Voter verifiability plays a key role in preventing
a compromised election authority from monopolizing real
credentials and issuing only fake ones to voters.

The Trust-limiting In-Person Voter Registration (TRIP)
system~\cite{TRIP}
addresses this problem by leveraging physical presence
of voters across four registration phases~(\cref{fig:TRIP:voter-workflow}):
check-in, credentialing, check-out, and activation.
At check-in,
voters identify themselves to a registration official,
obtaining a check-in ticket that gives them access
to a privacy booth for credentialing.
Comparable to the ``ballot selfie'' problem with in-person voting,
recording devices could compromise coercion resistance in this critical stage,
so TRIP assumes that voters cannot use electronic devices
in the booth.\footnote{
	The degree to which this rule is enforced is a policy decision,
	important but orthogonal to TRIP's design.
	Standard practice comparable to in-person voting would be
	merely to forbid the use of devices in the booth.
	Stronger enforcement might require voters
	to deposit electronic devices in a locker before entering the booth,
	at an obvious cost in convenience.}%
In the booth
voters find a kiosk, envelopes, and a pen.
Voters create their real credential in four steps:
\begin{enumerate}
    \item The voter scans their check-in ticket.
    \item The kiosk prints a QR code and symbol on receipt paper.
    \item The voter picks any envelope matching the printed symbol,
	  and presents the QR code on it to the kiosk's scanner.
    \item The kiosk prints two more QR codes on the receipt.
\end{enumerate}

These
steps establish an interactive zero-knowledge proof between the kiosk
and the voter, ensuring the correctness of the real credential.
The voter verifies that the kiosk adheres
to these steps while the voter's device---later, during activation---%
verifies the proof's cryptographic validity.

To finalize their credential,
the kiosk instructs voters to tear off the
receipt~(\cref{fig:TRIP:credentials:receipt}),
insert it into the envelope~(\cref{fig:TRIP:credentials:envelope}),
and memorably mark the resulting combination --
which we term the \emph{real paper credential} --
to distinguish it from future fake credentials.
When the receipt is fully inserted and
only its middle QR code is visible,
the credential is in its
\emph{transport state}~(\cref{fig:TRIP:credentials:transport}).

Voters may then create fake credentials in two steps:
\begin{enumerate}
    \item The voter chooses and scans any envelope.
    \item The kiosk then prints the receipt all at once.
\end{enumerate}

To finalize the fake credential,
voters again tear off and insert the receipt
inside the envelope,
marking the credential distinctively.
Voters may create any number of fake credentials,
limited only by their time in the privacy booth.\footnote{
	Whether to impose any particular time limit is a policy decision.
	We anticipate that voters spending an inordinate time registering
	should be a rare situation,
	manageable informally by an election official
	asking ``is anything wrong?'' at some point
	and gently escalating only as needed.
}

These two steps,
while also establishing an interactive
zero-knowledge proof, compromises the proof's soundness
compared to the four-step process.
Despite the proof's soundness being violated,
the proof's cryptographic validity remains unaffected.
Therefore, the voter's real credential
is cryptographically concealed among their fake credentials,
differentiated only by the voter's own distinct markings.

Upon exiting the booth,
voters proceed to check-out
where they present any one of their credentials, real or fake,
to the registration official.
The official scans the receipt's middle QR code,
visible through the envelope's transparent window, to
complete the in-person portion of the process.

Sometime later,
voters activate their real
credential on any device they trust by placing the credental in the
Activate state~(\cref{fig:TRIP:credentials:activate})
and scanning it.
After activation,
the voter discards the now-unusable paper credential.
Voters who have no device they trust
may activate their real credential on a device
of a trusted friend or family member.\footnote{%
	In hopefully-rare cases where a voter cannot
	hide a paper credential from the coercer,
	or have no access to any device they trust,
	TRIP offers an extension where
	voters can leave the booth with only fake credentials,
	while delegating their real vote to a designated proxy
	such as a political party.
	We do not explore this extension in the present study, however.}
Voters may give away or sell their fake paper credentials,
or activate them on a device that is under a coercer's control.

TRIP's design does not limit the lifetime of voting credentials.
Voters may therefore reuse their activated credentials
to cast (both real and fake) votes in multiple successive elections,
thereby amortizing the convenience cost of in-person registration.
Election authorities may impose an expiration date on voting credentials by policy,
perhaps aligning with the renewal cycle of
identification documents.

TRIP is designed to replicate the trust assumptions inherent in in-person
voting to achieve verifiability and coercion resistance.
In traditional in-person voting,
the election authority mitigates the risk of coercion by providing
voters with a privacy booth free from any coercer's influence.
This supervised environment creates an untappable communication
channel between voters and the election authority,
enabling voters to vote their conscience.
Voters, in return, collectively protect the integrity
of the election by reporting inconsistencies with their ballot
prior to submitting it~\cite{bernhard2020VotersDetectManipulation}.
TRIP, being designed for online voting,
leverages a similar untappable channel but
at registration time rather than voting time.
While a privacy booth traditionally hides a voter's choices,
in TRIP it hides knowledge of how many credentials the voter created
and which one is real.\footnote{
The design TRIP paper~\cite{TRIP} includes a formal proof that
coercers cannot distinguish between real and fake credentials,
or identify the number of fake credentials created by a voter
while inside the booth.}
Voters can therefore convey their intentions to the election authority
later while casting a vote online---%
by choosing to cast their vote using a real or fake credential.
TRIP relies on interactive zero-knowledge proofs
to ensure the integrity of the election process.
Voters need not understand zero-knowledge proofs,
but to verify the process's integrity,
they \emph{do} need to distinguish the four-step process
of creating a real credential
from the two-step process of creating a fake credential,
and to report any inconsistencies they encounter.

\subsection{Standardized Usability Metrics}
\noindent
In this study,
we use the System Usability Scale (SUS) and
the User Experience Questionnaire (UEQ) to measure
participants' perceptions of system usability and
user experience.

\parhead{System Usability Scale (SUS).}
This scale~\cite{
brooke1996SystemUsabilityScale}
is most often used to measure usability,
requiring only ten prompts. %
We altered the first prompt from
``I would like to use this system frequently'' to
``I would like to use this voter registration
system whenever I renew my identifications documents
(\ie every 5-10 years)'' to align with TRIP's intended usage.

\parhead{User Experience Questionnaire (UEQ).}
This questionnaire~\cite{laugwitz2008UEQ},
unlike SUS,
assesses not only traditional usability factors but
also aspects of user experience,
thereby ensuring the system meets user needs.
User experience covers a participant's emotional, cognitive,
and physical responses before, during, and
after system usage~\cite{hinderks2019UEQKPI}.
The UEQ consists of 26 contrasting attributes,
with participants expressing their level of agreement
by selecting a value between 1 and 7,
then re-calibrated to a score between -3 to 3.
The responses allow
for the measurement of six scales~\cite{laugwitz2008UEQ},
such as Perspicuity, Efficiency, Dependability and Novelty,
in addition to the individual scores obtained for each attribute,
such as security and practicality.
Furthermore,
Hinderks et al.~\cite{hinderks2019UEQKPI} introduce
a Key Performance Indicator (KPI) extension to the
UEQ, termed UEQ PKI.
This extension allows participants to express their
views on the significance of each of the six UEQ
scales via six additional questions. %

\section{Methodology}\label{sec:methods}
\noindent
This section presents our study's design,
detailing its objectives, 
workflow, materials and evaluation metrics.

\parhead{Ethical Considerations.}
Before participating,
each individual received an information sheet outlining
the study's scope, their rights, the expected tasks,
and the data to be collected.
We collected data from participants' interaction with
our devices and credentials, supplemented with
information from a survey.
We respected participants' privacy 
by supplying all required materials, 
and by not asking for any form of ID,
although showing ID would be standard
in real voter registration.
Furthermore,
we informed participants that this system
is not linked to any real voting system.

\subsection{Objectives}
\label{sec:methods:objectives}
\noindent
We start by detailing our study's central questions,
along with our strategies for addressing them.

\smallskip
\parhead{What are voters' perceptions and experiences
with coercion and how do they rate their
trust in this coercion-resistant 
voting system versus other voting methods?}
To answer the former,
we ask participants whether they have experienced or know
of someone who has experienced coercion.
The wording ``or someone you know'' is designed to
encourage disclosure from those who might otherwise
be hesitant to share their experience. 
We also ask to rate the likelihood of 
various coercion and vote-buying scenarios,
along with potential perpetrators.
For the latter,
we ask participants to rate the trustworthiness
of this coercion-resistant voting system in
comparison to other voting methods.

\smallskip
\parhead{Can voters use a credentialing process to create their
real and fake credentials, and identify deviations during
credentialing to ensure voter verifiability?}
To measure success rate,
we observe the number of participants who complete
the process successfully, without requiring 
a facilitator.
While during the survey,
we measure participants' experience and perceptions of
usability using both the SUS and UEQ questionnaires.
To measure the malicious kiosk detection rate,
we intentionally make the kiosk misbehave
and observe if participants report the anomaly.

\smallskip
\parhead{Can voters understand and use fake credentials, while still
casting their intended votes using their real credential?}
We gather data about participants' understanding
of fake credentials in two phases:
first via an interactive quiz on the kiosk,
then later via survey questions.
We also present an instructional video at the beginning of the process
to prepare and educate the voter.
Since real and fake credentials function identically,
we direct participants to vote using their real credential,
allowing us to assess whether they can distinguish
their real credential from their fake credentials.

\subsection{Study Setup \& Workflow}
\label{sec:study:setup}
\noindent
We now describe the study's setup and workflow,
detailing its location, the setup of the workstation 
and the participant workflow.
The facilitator's scripts are
available in~\cref{apx:study:materials:scripts}.

\smallskip
\parhead{Location and Recruitment.}
\begin{figure}[t]
    \centering
    \includegraphics[width=0.9\linewidth]{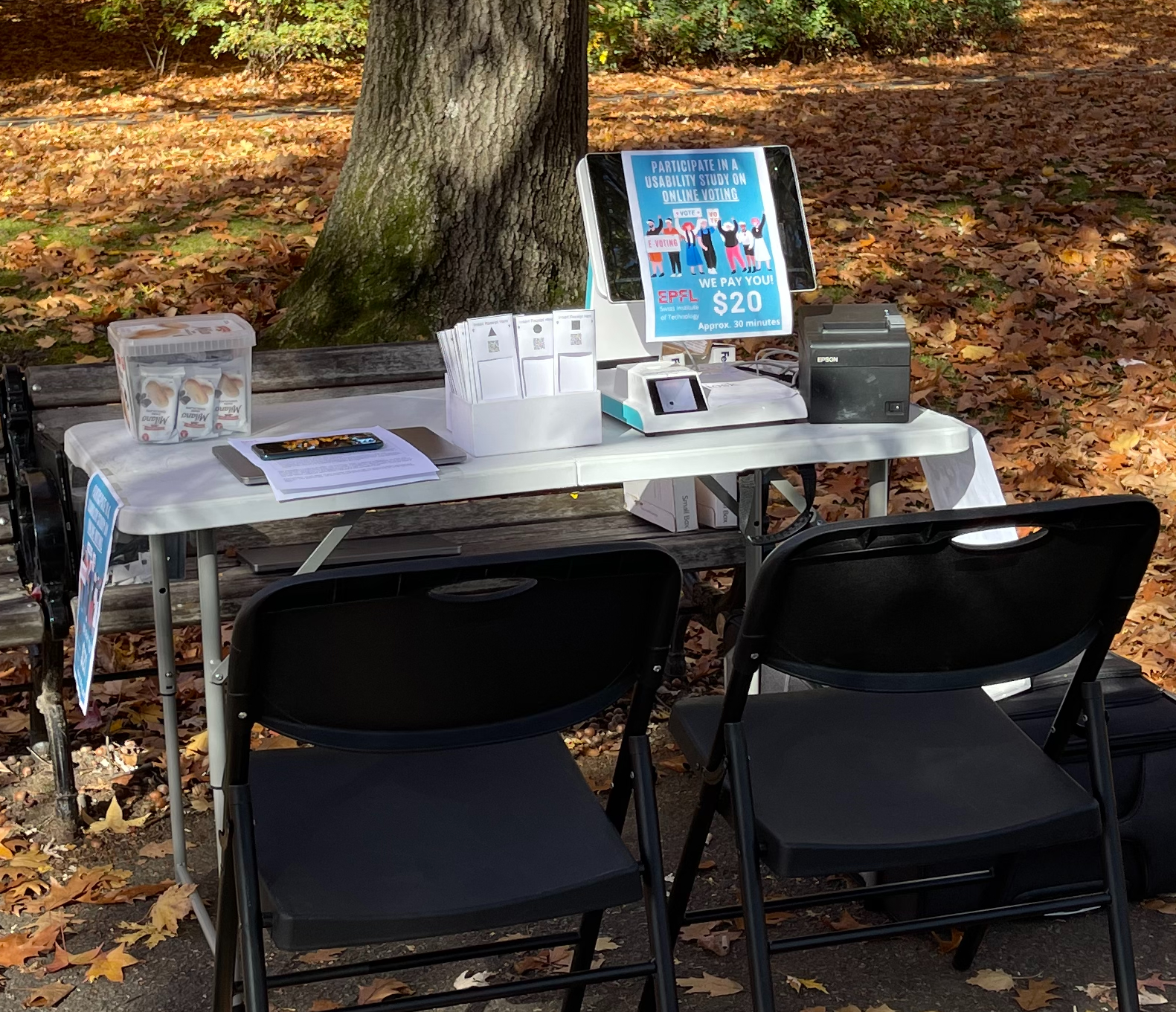}
    \caption{
    \textbf{Study Setup}:
    Starting on the left chair, participants 
    (1)  review the study's information sheet and sign the consent form,
    (2) watch an instructional video on the laptop,
    (3) are handed a check-in ticket from the facilitator,
    (4) move to the right chair and 
        interact with the kiosk to create their credential(s),
    (5) return to the left chair for check-out,
    (6) activate their credential(s) and cast
        mock votes on the supplied mobile device, and,
    (7) complete the study's exit survey on either 
        our laptop or touchscreen tablet.
    }
    \label{fig:study:setup}
\end{figure}

To enhance demographic diversity
while ensuring a neutral environment,
we conduct our study in a suburban park.
We verbally invited passerbys to participate in the study 
over three months, and showcased flyers/posters at our 
study location to encourage engagement.
We limited participants to only those who said they had 
prior experience with voter registration.

\smallskip
\parhead{Study Setup.}
Our setup (\cref{fig:study:setup})
consists of a table with 
two chairs on one side for the participant
and a park bench on the other side for the facilitator.
On the right side of the table, 
we have the credentialing process with the kiosk, 
envelopes and receipt printer,
while on the left side,
we have the check-in, check-out, activation
and voting processes.

\smallskip
\parhead{Participant Enrollment.}
Whenever an eligible participant agrees to participate,
we guide them to the left side of the table
to review the information sheet and sign the
consent form.
The facilitator addresses any logistical questions, 
and refrains from discussing the objectives of 
the study beyond ``assessing the usability of
an online voting system.''
The participant is then randomly assigned to one
of five groups that we detail
in~\cref{sec:study:groups}.
To enhance the ecological validity of our study~\cite{
kjeldskov2007StudyingUsabilitySitro},
the facilitator asks the participant to envision 
themselves at a government office, as scripted 
in~\cref{apx:study:materials:scripts}.

\parhead{Instructional Video \& Check-In.}
The facilitator then starts playing a video (\cref{sec:study:videos})
for the participant to watch,
the content of which depends on the participant's assigned group.
After the video concludes, 
the facilitator hands the participant a check-in ticket 
and directs them to the booth.

\parhead{Credentialing and Check-Out.}
To simulate an authentic booth experience,
the facilitator does not interact
with the participant in this stage, intervening only
upon request.
Upon completing the credentialing process,
the participant returns to the left side
of the table and hands one credential
(real or fake) to the facilitator.
The facilitator scans the QR code
through the visible window and
returns the credential.

\parhead{Activation and Voting.}
For activation,
the facilitator asks the participant to envision that they
are now at home, and to use the study's supplied mobile
device to activate their real credential and cast a
mock vote.
The participant may optionally activate fake
credential(s) as well and cast fake votes by clicking
on `cast vote with another credential'.
Since the app was designed solely for the purpose of this study,
it is stateless,
storing only one credential at a time.
If asked, we explain
that in a real setting, one could 
unlock a specific credential using its associated user-generated PIN.

\parhead{Survey and Compensation.}
Once finished,
the participant completes the exit survey
and is then compensated with \$20.
The entire session typically takes around
30 minutes.

\subsection{Study Groups}
\label{sec:study:groups}
\begin{figure}[t]
    \centering
    \includegraphics[width=\linewidth]{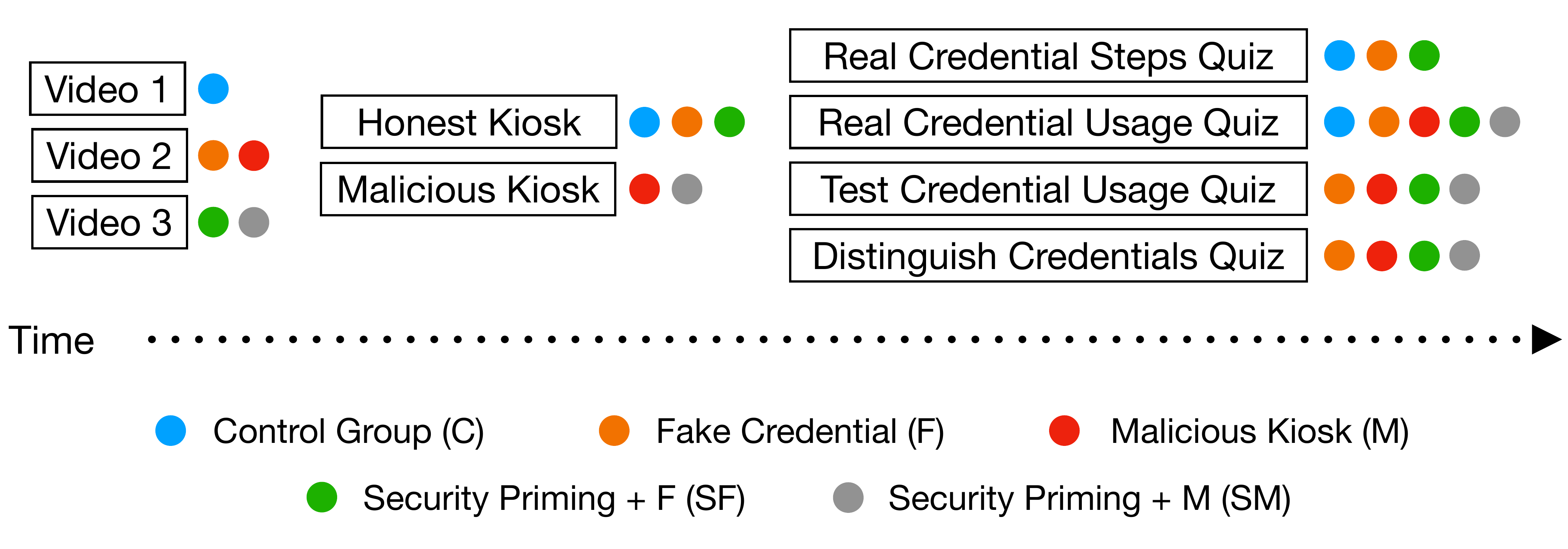}
    \caption{\textbf{Study Flow.}
        Participants first watch one of three instructional videos,
        then interact with either an honest or malicious kiosk
        that displays the corresponding quizzes.
        The distinguish credentials quiz is present for those
        who created at least one fake credential.
        }
        \label{fig:study:flow}
\end{figure}

\noindent
To answer our key questions~(\cref{sec:methods:objectives}),
we conduct a between-subjects study,
randomly assigning study participants
to five equal-size groups.
Each group experiences a distinct variant of the process.
To assess the usability of fake credentials in particular,
our study exposes one control group only to real credentials,
while exposing the other four groups to real and fake credentials.
To assess the system's effective voter verifiability,
we expose two of the latter groups to
a ``malicious'' kiosk that attempts to ``steal'' the voter's
real credential by silently guiding the user through the creation of a fake
credential instead of a real one.\footnote{
The deviation involves reordering steps 2 and 3, with step 3 
now being before 2.
Other deviations are possible 
and perhaps worth studying in the future,
but most variations are either more overt and obvious
(\eg skipping the creation of a real credential) 
or detectable by the voter's personal device on activation 
(\eg the cryptographic validity of the zero-knowledge
proof).}
To familiarize participants with the correct 
voting process we rely on instructional videos,
as detailed in~\cref{sec:study:videos}.
To study the tradeoffs between being more or less explicit
about security threats and risks in educational materials,
we use two contrasting types of instructional videos.
We therefore obtain the following five groups,
each consisting of 30 participants:
\begin{itemize}
    \item \textbf{Control Group (C)}: Exposure only to the real
    credential known as ``voting credential'' and
    instructional video 1.
    \item \textbf{Fake Credential Group (F)}: 
    Exposure to real and fake voting credentials, and 
    instructional video 2.
    \item \textbf{Malicious Kiosk Group (M)}: 
    Exposure to real and fake credentials, 
    instructional video 2, and a malicious kiosk.
    \item \textbf{Security Priming + F (SF)}: Exposure to 
    real and fake voting credentials, and
    instructional video 3.
    \item \textbf{Security Priming + M (SM)}: 
    Exposure to real and fake credentials,
    instructional video 3, and a malicious kiosk.
\end{itemize}

\subsection{Materials}
\label{sec:study:videos}
\begin{figure}[t]
    \centering
    \includegraphics[width=\linewidth]{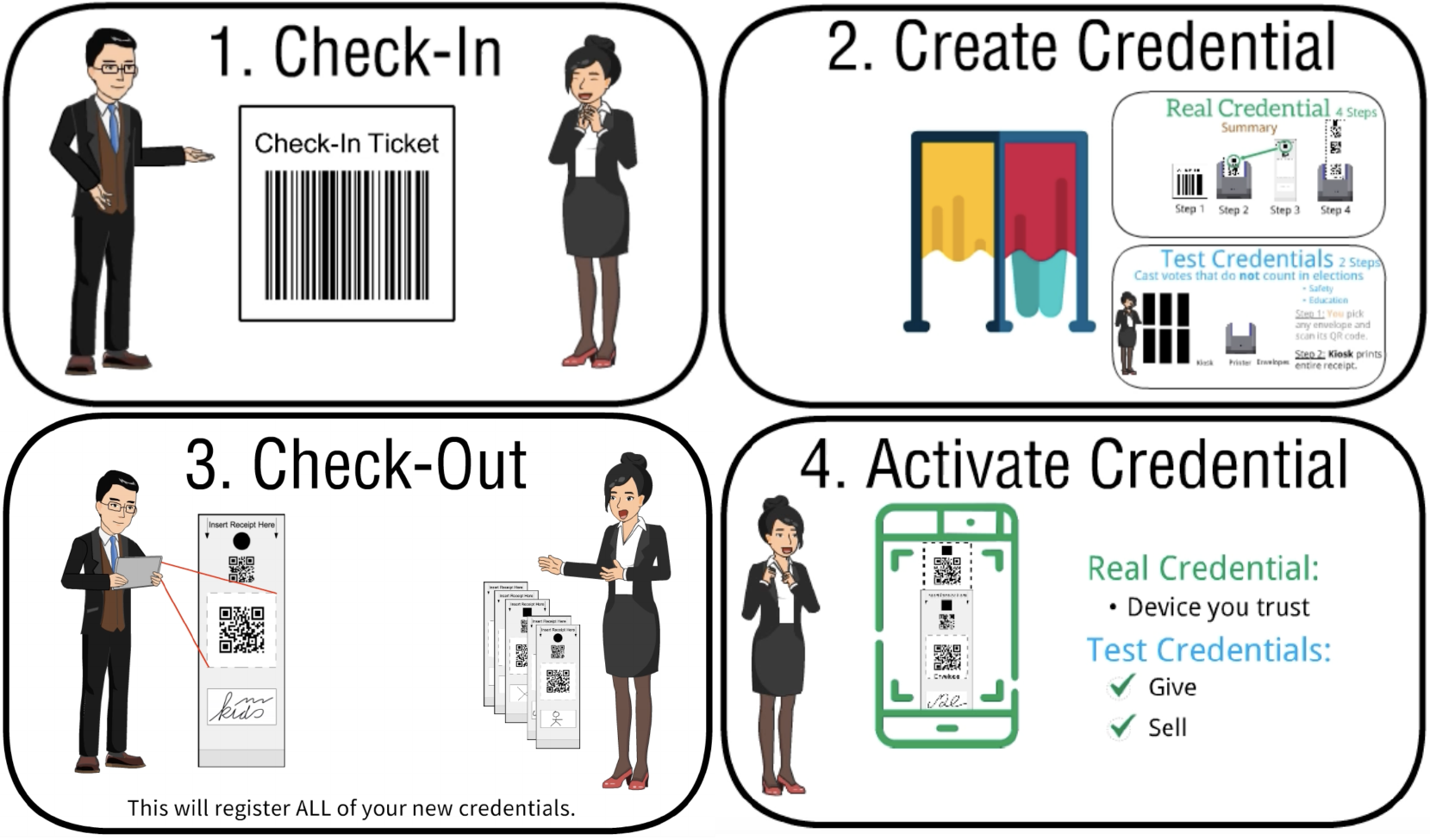}
    \caption{\textbf{Steps Overview.}
        Various screenshots from the instructional video, which demonstrate
        each phase.}
    \label{fig:video:overview}
\end{figure}

\noindent
This study uses instructional videos to educate voters,
while quizzes and a survey assist us in evaluating their 
understanding according to the study flow presented in Figure~\ref{fig:study:flow}.

\parhead{Instructional Videos.}
We introduce three videos, available at
\href{\materialslink}{github.com/dedis/trip-usability},
each illustrating their assigned conditions: control group (video 1), 
no security priming (video 2) and security
priming (video 3).
Video 1, lasting 3 minutes and 6 seconds,
demonstrates only the creation of a ``voting'' credential.
In contrast, videos 2 and 3, with durations of 5m 50s and 6m 25s respectively,
illustrates the creation of real and fake credentials (\cref{fig:video:overview}).
Videos 2 (\cref{fig:video:b:creds}) 
and 3 (\cref{fig:video:c:creds})
both highlight the differences in creating 
real and fake credentials.
However, video 3 shows, under 
a conspicuous ``BEWARE'' sign,
how to detect a ``hacked'' kiosk.
Video 3 thus makes a key threat explicit,
at the risk of a more unsettling or ``scary'' presentation.

We use videos as our primary source of instructional
material, as individuals learn better %
from dynamic visuals compared to static images~\cite{
ploetzner2021AnimationsVsStaticImages}.
We select whiteboard animated videos due
to their positive impacts on retention, engagement
and enjoyment, even when conveying complex material~\cite{
turkay2016WhiteboardAnimations}.
To enhance the effectiveness of our videos,
we incorporate the dynamic drawing principle,
adopt a first-person perspective, and 
include narration with subtitles~\cite{
mayer2020VideoLearningEffectiveness}.

\parhead{Quizzes.}
\label{sec:methods:quizzes}
To evaluate participants' understanding of concepts
and foster active learning,
the kiosk exposes participants to up to four
unannounced, multiple-choice quizzes (\cref{apx:study:quizzes}.
The selection of quizzes depends on the participant's
group assignment and actions.
For all groups, 
we evaluate participants' understanding that they must
keep and activate their real credential on a trusted device.
For the treatment groups (F, M, SF, SM),
we examine participants' recall of the stated purposes
of fake credentials.
For groups C, F, and SF,
we also assess comprehension of when to select and scan
an envelope prior to creating their real credential.
Finally,
for those electing to create fake credential(s),
we evaluate their recall on how to differentiate
their real credential from fake credentials.

\parhead{Exit Survey.}
In the exit survey
(\cref{apx:survey}),
we first ask participants their
demographic information.
Participants then rate their experience,
including what they liked most and liked least,
along with completing the SUS, UEQ, and UEQ PKI 
questionnaires.
Participants continue by answering whether
they noticed anything odd with the 
credentialing process.
Participants exposed to fake
credentials answer questions
related to those credentials, such as
whether they can recall their usage.
Participants then rate their 
trust on various voting methods:
three variants of in-person voting,
and three variants of remote voting,
including one for this voting system.
The survey finishes by asking participants
to describe their
perceptions and experiences with coercion
and vote-buying in their own lives.
The average time spent on the survey is 17 minutes.

\subsection{Statistical Methods}
\noindent
Throughout this paper,
we use an alpha level of $0.05$ to 
establish statistical significance.
For our between-subjects comparison 
on non-parametric data (\eg ordinals),
we use the Kruskal-Wallis one-way ANOVA,
while for pairwise comparisons, we use
the Dunn test.
For our within-subjects comparisons
on non-parametric data,
we use the Friedman rank sum test,
while for pairwise comparisons,
we use the Durbin-Conover test.
To control for the family-wise error rate when conducting
pairwise statistical comparisons,
we use the Holm-Bonferroni method.
We apply Shapiro-Wilk to assess whether 
the current data follows a normal distribution.
We employ Cohen's $d$ to measure effect sizes,
which estimate the degree of difference between two
groups.
Typically, an effect size of about 0.2 is considered
small, while an effect size near 0.8 is considered large.
The term ``participants'' represent
the average views or experiences of those involved in the study.

\section{Participants' Perceptions and Experiences}
\label{sec:participants}
\noindent
This section details the demographic profile of our participants 
and discusses their experiences with and perceptions of coercion. 
Moreover, we explore their views on this coercion-resistant system
versus other voting methods.

\subsection{Demographics}
\begin{figure}
    \centering
        \centering
         \includegraphics[width=0.88\linewidth]{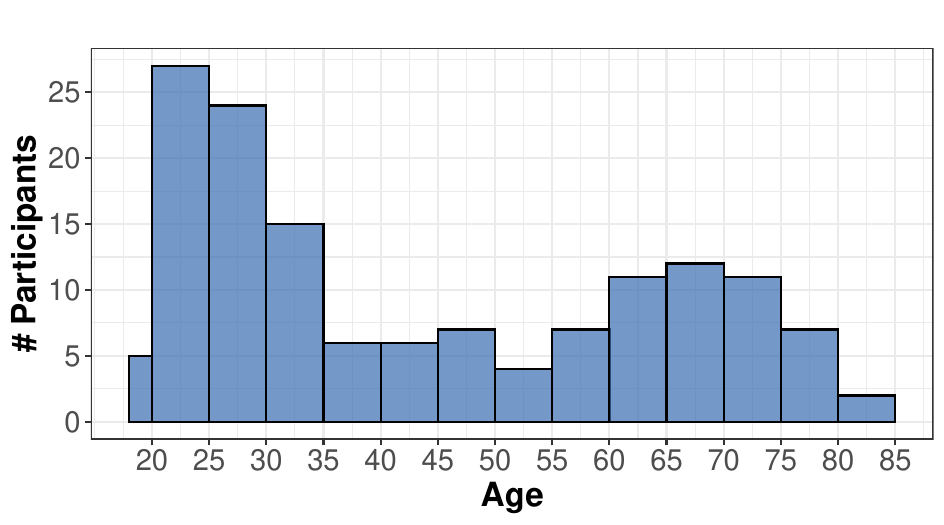}
         \caption{\textbf{Distribution of Participant Ages}.
            The minimum, median, mean and maximum ages among
            all groups were 19, 36.5, 44 and 83, respectively.
            Six participants did not disclose their age.}
         \label{fig:age:simple}
\end{figure}

\begin{figure*}[!t]
    \centering
    \hfill
    \begin{subfigure}[b]{0.39\textwidth}
        \centering
        \includegraphics[width=\textwidth]{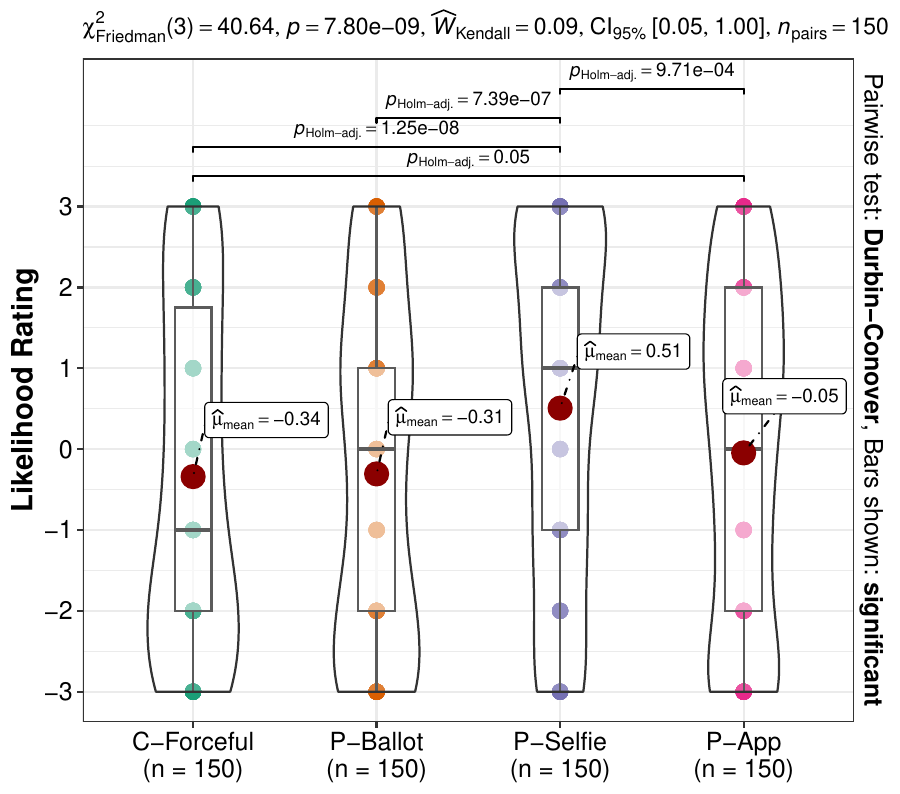}
        \caption{\textbf{Coercion Scenarios.} 
        Participants perception of likely coercion scenarios;
        ballot selfies received the highest average likelihood,
        while forceful coercion received the lowest.}
        \label{fig:coercion:scenarios}
    \end{subfigure}
    \hfill
    \begin{subfigure}[b]{0.39\textwidth}
        \centering
        \includegraphics[width=\textwidth]{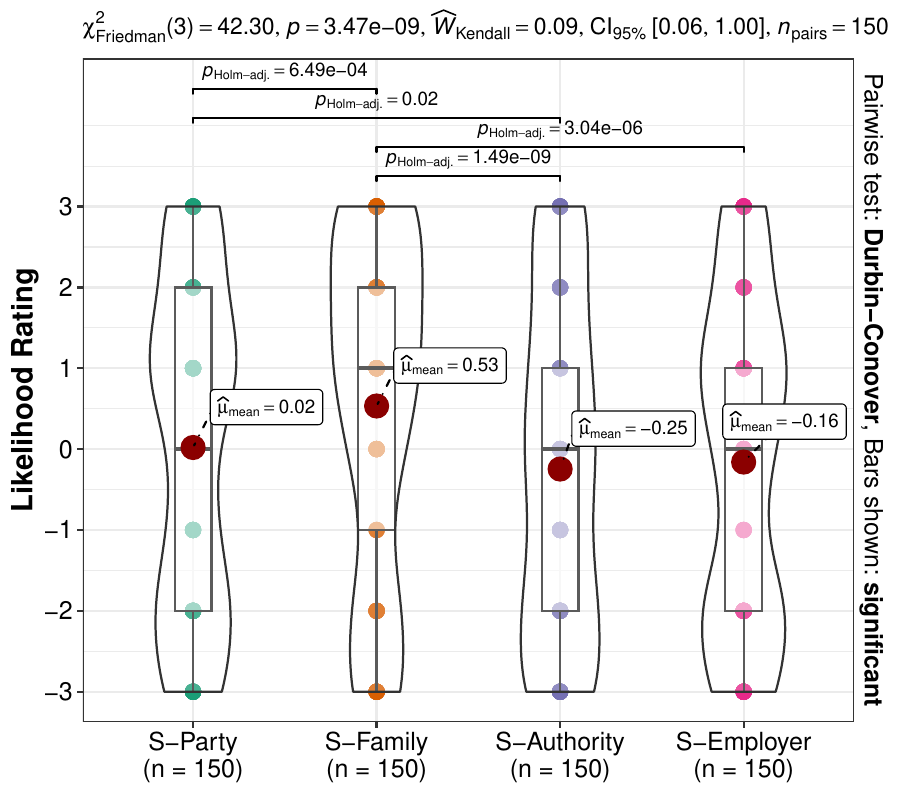}
        \caption{\textbf{Coercion Sources.} Participants perception of likely sources of coercion;
        family members received the highest average likelihood rating, while authority figures
        received the lowest.}
        \label{fig:coercion:sources}
    \end{subfigure}
    \hfill
    \begin{subfigure}[b]{0.19\textwidth}
        \centering
        \scriptsize
        \begin{tabularx}{\linewidth}{
            >{\centering\arraybackslash\hsize=2.1\hsize}X
            >{\raggedleft\arraybackslash\hsize=0.3\hsize}X
            >{\raggedleft\arraybackslash\hsize=0.6\hsize}X}
            \toprule
            \textbf{Scenario} & Hap-pened & Not Happened\\
            \midrule
            C-Forceful          & 12\% & 81\% \\
            P-Ballot            & 8\%  & 85\% \\
            P-Selfie            & 6\%  & 86\% \\
            P-App               & 3\%  & 88\% \\
            Total Count         & 44    & 511    \\
            Unique Count        & 26    & 112    \\
            \midrule
            \textbf{Source} & 
                \multicolumn{2}{c}{} \\[1ex]
            S-Party             & 10\% & 81\% \\
            S-Family            & 15\% & 77\% \\
            S-Authority         & 7\%  & 85\% \\
            S-Employer          & 11\% & 83\% \\
            Total Count         & 63    & 491    \\
            Unique Count        & 31    & 108    \\
            \midrule
            Unique Participants & 39 & 98 \\
            \bottomrule
        \end{tabularx}
        \caption{\textbf{Coercion Instances.} 
        Participants state whether
        these items have happened to them or someone they know.
        The remaining participants chose not to disclose.}
        \label{fig:coercion:occurrence}
    \end{subfigure}
    \caption{\textbf{Participants' Views and Experiences with Coercion Scenarios, and Coercion Sources}}
    \label{fig:coercion}
\end{figure*}

\noindent
We recruited 150 participants, aged 19 to 83,
with an average and median age of 44 and 36.5,
respectively.
\Cref{fig:age:simple} depicts this age distribution.
\Cref{fig:age} provides a breakdown of the 
participants' age distribution across the five study groups,
while \Cref{fig:categorical-demographics} provides 
a breakdown of their gender, ethnicity and education.
We observed distinct participation patterns:
seniors primarily during the day,
middle-aged individuals after work,
and younger individuals (18-35) throughout the day. 
On initial encounter,
many individuals had the first impression that they could
register with us for real online voting; we clarified in our
recruitment script (\cref{apx:study:materials:scripts})
that this is only a usability study on 
online voting with mock, non-political, elections.
Participants' time availability was the primary recruitment challenge;
we estimated the study to last around 30 minutes, and found, after the fact,
that a typical participant took 35 minutes.
A few individuals declined to 
participate due to their opposition
to online voting, reconfirming well-known difficulties
in overcoming selection bias in user studies.

\subsection{Coercion and Vote Buying}
\noindent
We now discuss
participants' perspectives and experiences 
with coercion and vote buying
based on their survey responses. %
Participants rated the following four coercion scenarios 
on a 7-point Likert scale~\cite{likert1932LikertScale} 
(summarized from \cref{apx:survey:coercion-scenario}):
(\texttt{C})oercing by threatening harm (\texttt{C-Forceful}),
(\texttt{P})urchasing absentee ballots (\texttt{P-Ballot})\footnote{
Illustrated by the 
North Carolina ballot fraud incident~\cite{2022NCAbsenteeFraud}.},
purchasing proof such as a voter taking a selfie with their 
ballot containing the coercer-dictated choices (\texttt{P-Selfie}),
and an app offering compensation to vote as directed
(\texttt{P-App}).
They also rated the following four potential sources of coercion
on the same scale:
a party operative (\texttt{S-Party}),
an authority figure (\texttt{S-Authority}),
a family member (\texttt{S-Family}),
and an employer (\texttt{S-Employer}).
Ratings are then converted to a score
between -3 to 3, where -3 is `completely unlikely'
and 3 is `completely likely'.

\parhead{Perceived Scenarios (\cref{fig:coercion:scenarios}).}
Among the coercion scenarios,
participants find \texttt{P-Selfie}
to be the most likely scenario 
with a mean score of 0.51 and a median of 1 (somewhat likely).
A quarter (24\%) of participants rated \texttt{P-Selfie} 
as extremely likely, with this scenario being perceived as
having a statistically significantly higher likelihood
compared to the other three coercion scenarios.
In contrast,
participants perceived \texttt{C-Forceful} 
as the least likely scenario with 
a mean score of -0.34 and a median of -1 (somewhat unlikely).
We also see that the 
distribution of responses for \texttt{C-Forceful}
almost inversely mirrors that of \texttt{P-Selfie}.
The calculated Cohen's $d$ effect sizes further quantify these differences:
0.41 when comparing \texttt{P-Selfie} with \texttt{C-Forceful}, 
0.39 with \texttt{P-Ballot} and 0.26 with \texttt{P-App}.
The remaining two coercion scenarios, 
\texttt{P-Ballot} and \texttt{P-App}
also bear a visual similarity in their distributions,
demonstrating two distinct peaks at somewhat likely
and extremely unlikely.

\parhead{Perceived Sources (\cref{fig:coercion:sources}).}
Participants perceive \texttt{S-Family} 
to be the most likely source of coercion
(mean score of 0.53 and a median rating of somewhat likely),
with 21\% of participants rating it as extremely 
likely.
Moreover,
\texttt{S-Family} is statistically significantly higher
than the three other coercion sources.
In contrast, \texttt{S-Authority} is the least likely
source with a mean score of -0.25.
The calculated Cohen’s $d$ effect sizes are: 
	0.38 when comparing \texttt{S-Family} with \texttt{S-Authority}, 
	0.34 with \texttt{S-Employer}, and
	0.25 with \texttt{S-Party}.
For both \texttt{S-Employer} and \texttt{S-Party} sources,
participants have two dominant contrasting opinions,
with one cohort seeing the source as somewhat likely
and the other as very unlikely.

\parhead{Coercion Instances (\cref{fig:coercion:occurrence}).}
A quarter of participants (26\%) report experiencing
or knowing of someone who has experienced at least one form
of voter coercion,
while two-thirds (67\%) report no such experiences; 
the remaining 7\% preferred not to say.
In line with participants' views, 
the most commonly reported source of coercion (15\%)
is \texttt{S-Family}.
Such instances include multiple accounts of
spousal oversight during voting, leading to
dictated voting choices.
However, despite being viewed as the least likely coercion scenario,
\texttt{S-Forceful} is the most reported coercion scenario (12\%).
Around 10\% identify \texttt{S-Employer}
and \texttt{S-Party} as sources of coercion.
Reported incidents include unions dictating votes,
co-workers pressuring attendance at undesired political rallies,
and prominent party members pressuring members to vote along party lines.

\parhead{Discussion.}
The substantial reporting rate of recognized coercion instances at 26\%, 
with family members cited as the predominant coercion source (15\%), 
highlights the need for coercion-resistant strategies to counter 
\emph{persistent} adversarial oversight.
In such circumstances,
the practice of deniable re-voting,
such as the one in use in Estonia~\cite{kulyk2020HumanFactorsCoercion},
falls short.
Family members, typically having substantial unrestricted access, 
can potentially cast a vote on behalf of the targeted relative 
just before the election concludes. 
Such a coercer can also realistically stay with the relative or retain 
their device or voting credential until the 
voting process ends.

\subsection{Trust in Voting Methods}
\begin{table}[t]
    \centering
    \footnotesize
    \begin{tabularx}{\linewidth}{
    >{\centering\arraybackslash\hsize=2.8\hsize}X
    >{\raggedleft\arraybackslash\hsize=0.7\hsize}X
    >{\raggedleft\arraybackslash\hsize=0.7\hsize}X
    >{\raggedleft\arraybackslash\hsize=0.7\hsize}X
    >{\raggedleft\arraybackslash\hsize=0.7\hsize}X
    >{\raggedleft\arraybackslash\hsize=0.7\hsize}X
    >{\raggedleft\arraybackslash\hsize=0.7\hsize}X}
        \toprule
        \textbf{Voting Method} & I-Ballot & I-BMD & I-DRE & R-Mail & R-Online & IR-Online \\
        \midrule
        Mean          &   1.38      &   1.68   &   1.18   &   0.72    &   0.82      &  1.25   \\
        Median        &   2         &   2      &   1.5      &   1       &   1         &  2   \\
        \midrule
        \textbf{Trust Rating} & 
            \multicolumn{6}{c}{\textbf{\textit{Participants (\%)}}} \\[1ex]
        Trusted (3)             &   35    & 36    & 26       & 22       & 20        & 20   \\
        Highly                  &   25    & 33    & 24       & 24       & 22        & 35   \\
        Somewhat                &   12    & 11    & 18       & 15       & 17        & 16   \\
        Neutral     (0)         &   12    & 10    & 17       & 12       & 18        & 17   \\
        Somewhat                &   7     & 5     & 7        & 7        & 9         & 5   \\
        Highly                  &   6     & 3     & 4        & 12       & 6         & 4   \\
        Untrusted (-3)          &   3     & 3     & 3        & 8        & 7         & 3  \\
        \midrule
        \textbf{Voting Method} & 
            \multicolumn{6}{c}{\textbf{\textit{Pairwise Statistical Outcomes (\cref{fig:voting-methods:fake-credentials})}}} \\[1ex]
        I-Ballot   &                & $\times$     & $\times$     &  $\uparrow$ & $\uparrow$   &  $\times$      \\
        I-BMD      &  $\times$      &              & $\uparrow$   &  $\uparrow$ & $\uparrow$   &  $\uparrow$    \\
        I-DRE      &  $\times$      & $\downarrow$ &              &   $\times$  & $\times$   &  $\times$      \\
        R-Mail     &  $\downarrow$  & $\downarrow$ & $\times$     &             & $\times$     &  $\times$      \\
        R-Online   &  $\downarrow$  & $\downarrow$ & $\times$ &   $\times$  &              &  $\times$  \\
        IR-Online     &   $\times$  & $\downarrow$ & $\times$     &   $\times$  & $\times$   &                \\
        \bottomrule
    \end{tabularx}
    \caption{\textbf{Voting Methods.}
    This table presents the trust ratings for various voting methods,
    as given by the 120 participants exposed to fake credentials,
    and a summary of the statistical outcomes~(\cref{fig:voting-methods:fake-credentials}).
    $\times$ means no statistical difference, $\downarrow$ means statistically significantly lower,
    and $\uparrow$ means statistically significantly higher.}
    \label{tab:voting-methods}
\end{table}

We now investigate the level of trust 
the 120 participants in treatment groups
perceive in our coercion-resistant system
versus other voting methods.
Participants rated their level of trust in the
following voting methods on a 7 point Likert Scale (summarized from~\cref{apx:survey:voting-methods}):
(\texttt{I})n-person voting with hand-marked paper ballot (\texttt{I-Ballot}),
in-person voting with a ballot marking device (\texttt{I-BMD}),
in-person voting with a direct electronic device (\texttt{I-DRE}),
(\texttt{R})emote voting via mail-in ballot (\texttt{R-Mail}),
a fully-remote voting system where both voter registration
and voting are online (\texttt{R-Online}),
and the online voting system they just experienced with
in-person voter registration (\texttt{IR-Online}).
\Cref{tab:voting-methods} presents our summarized results 
while \Cref{fig:voting-methods} in Appendix~\ref{apx:misc}
contains more complete statistical results.

\parhead{Summarized Results.}
Analyzing the trust scores 
and the pairwise statistical outcomes across voting methods,
we identify three cohorts.
Participants place the highest trust in 
\texttt{I-BMD} and \texttt{I-Ballot},
and least in \texttt{R-Mail} and \texttt{R-Online},
with \texttt{I-DRE} and \texttt{IR-Online}
in-between.

\parhead{IR-Online.}
Participants generally regard \texttt{IR-Online} 
as `somewhat trustworthy', as indicated by a mean rating of 1.25. 
However, the median participant sees it as `highly trustworthy', 
as indicated by a greater median rating of 2.
This ranks \texttt{IR-Online} marginally higher 
than \texttt{I-DRE}, having a mean trust rating of 1.18, 
yet remains below \texttt{I-Ballot}, exhibiting a higher 
average of 1.38.
Moreover,
although we find no statistically significant
difference in trust ratings between the treatment groups,
the median trustworthiness rating for the security priming 
groups SF and SM is `somewhat trustworthy',
contrasting with groups F and M's `highly trustworthy' rating
(\cref{fig:voting-methods:ir-online}).
We also find no statistically significant difference in trust ratings
for \texttt{IR-Online} and \texttt{R-Online} between
the control group $(n = 30)$ and the treatment groups
$(n = 120)$ (Mann-Whitney U test: $p = 0.969$). %

\parhead{Participants Experiencing Coercion.}
We examine the trust ratings 
from the 33 participants (22\%) not in group C
who report personal or known experiences of coercion.
These results, depicted 
in~\Cref{fig:voting-methods:coercion-happened},
show statistically significant lower trust scores for
\texttt{R-Mail} and \texttt{R-Online} when compared to 
\texttt{I-BMD}, indicating a general distrust towards
remote voting methods.
Based on Cohen's $d$, the effect sizes 
between \texttt{I-BMD} and \texttt{R-Mail},
as well as \texttt{I-BMD} and \texttt{R-Online}, 
are 0.76 and 0.69 respectively.
Simultaneously, the score of \texttt{I-Ballot} dropped
to match \texttt{IR-Online}, while that of \texttt{I-DRE}
rose to match \texttt{I-BMD}.
This could possibly be influenced by the 
perceived risk of ballot selfies.

\parhead{Discussion.}
These findings demonstrate the system's 
promising potential to attain trustworthiness 
scores on par with in-person voting.
Presently,
the system's overall score exceed those of
in-person voting with a direct recording device,
employed in 11.5\% of U.S. jurisdictions in 2020~\cite{VerifiedVoting},
and approaches the trust levels associated with in-person
hand-marked paper ballots,
used in 68.1\% of jurisdictions.

\parhead{Additional Findings.}
In~\Cref{apx:misc:voting-methods},
we discuss our findings when we contrast
the other voting methods with each other,
particularly the unanticipated lower trust 
for \texttt{I-Ballot} compared to \texttt{I-BMD}.
We also discuss the
surprisingly low trust rating for \texttt{R-Mail},
despite its use by 46\% of voters in the 2020
U.S. presidential election~\cite{
atske2020VotingMethods2020Presidential}, and
the high trust rating in \texttt{I-BMD},
despite only 6.6\% of participants capable
of detecting a \emph{manipulated} BMD-printed
ballot~\cite{bernhard2020VotersDetectManipulation}.
Moreover,
we present our findings when considering all 150 participants,
particularly the statistically significantly higher trust
in \texttt{IR-Online} compared to \texttt{R-Online}.

\section{Usability of Registration and Voting}
\label{sec:usability}

\noindent
In this section,
we present the usability results for TRIP,
a voter-verifiable registration process
that outputs real and fake credentials and
relies on voters to identify deviations 
during credentialing to meet voter verifiability.
We begin with qualitative results,
discussing the system aspects participants
most liked and disliked.
We then present the observed use errors
and the number of 
participants who successfully 
completed the registration and voting 
process without facilitator intervention.
We continue with the System Usability Scale and
User Experience Questionnaire scores,
contrasting these against benchmarks and other
voting-related studies.
We conclude with our findings concerning 
the participants who encountered 
our purposely designed malicious kiosk.

\parhead{Groups.}
The control group (C) establishes the baseline
against which we evaluate the four experimental
groups (F, M, SF, and SM).
Group F serves as the expected norm,
with group M being relevant in the
event of a malicious kiosk.

\subsection{Qualitative Results}
\begin{table}[t]
    \centering
    \begin{tabular}{c c c|c c c}
    \toprule
        easy & positive & 17 & new & neutral & 3 \\
        interesting & positive & 13 & streamlined & positive & 3 \\
        complicated & negative & 8  & long & negative & 2 \\
        good & positive & 7 & fine & neutral & 2 \\
        simple & positive & 7 & fast & positive & 2 \\
        straightforward & positive & 4 & complex & negative & 2 \\
        great & positive & 4 & convenient & positive & 2 \\
        smooth & positive & 4 & cumbersome & negative & 2 \\
        confusing & negative & 4 & efficient & positive & 2 \\
    \bottomrule
    \end{tabular}
    \caption{
    \textbf{Distribution of Single-Word Summaries with Sentiment.}
    This table lists the words used more than once by participants to describe their impressions of the system along with each word's associated sentiment (positive, neutral or negative).
    }
    \label{tab:single-word}
\end{table}

\renewcommand\tabularxcolumn[1]{m{#1}}

\newcommand\upuparrow{\uparrow \uparrow}
\newcommand\downdownarrow{\downarrow \downarrow}

\begin{table}[!t]
\scriptsize
\begin{tabularx}{\linewidth}{
    >{\centering\arraybackslash\hsize=1.96\hsize}X
    >{\raggedleft\arraybackslash\hsize=0.78\hsize}X
    >{\raggedleft\arraybackslash\hsize=0.78\hsize}X
    >{\raggedleft\arraybackslash\hsize=0.9\hsize}X
    >{\raggedleft\arraybackslash\hsize=0.78\hsize}X
    >{\raggedleft\arraybackslash\hsize=0.78\hsize}X}
    \toprule
    \textbf{Groups} & 
      (C)ontrol & (F)ake Creds. & (M)alicious Kiosk & SF & SM \\
    \midrule
    \textbf{Rating} & 
      \multicolumn{5}{c}{\textbf{\textit{System Approval Rating}}} \\[1ex]
    Positive & 
    73\% & 63\% & 60\% & 56\% & 53\% \\
    Neutral & 
    13\% & 20\% & 3\% & 17\% & 23\% \\
    Negative & 
    13\% & 17\% & 37\% & 27\% & 23\% \\
    \midrule
    & 
      \multicolumn{5}{c}{\textbf{\textit{System Usability Scale}}} \\[1ex]
    Score & 69.6 & 70.4 & 69.9 & 67.3 & 62.7 \\
    SD & 18.6 & 18.6 & 17.4 & 19.8 & 21.9 \\
    N & 29 & 28 & 29 & 30 & 29 \\
    Percentile & 55.1 & 57.8 & 56.1 & 47.8 & 35.0 \\
    Usability & 69.5 & 69.8 & 68.4 & 67.7 & 62.5 \\
    Learnability & 69.8 & 73.2 & 75.9 & 65.8 & 63.4 \\
    \midrule
    \textbf{Scale} & 
      \multicolumn{5}{c}{\textbf{\textit{User Experience Questionnaire (Score vs. Benchmark)}}} \\[1ex]
    Attractiveness & 
    $\uparrow$ & $\uparrow$ & $\downarrow$ & $\downarrow$ & $\downdownarrow$\\
    Perspicuity & 
    $\uparrow$ & $\downarrow$ & $\downarrow$ & $\downarrow$ & $\downdownarrow$ \\
    Efficiency & 
    $\upuparrow$ & $\upuparrow$ & $\uparrow$ & $\downarrow$ & $\downarrow$ \\
    Dependability & 
    $\downarrow$ & $\uparrow$ & $\uparrow$ & $\downarrow$ & $\downdownarrow$ \\
    Stimulation & 
    $\upuparrow$ & $\upuparrow$ & $\uparrow$ & $\uparrow$ & $\downarrow$ \\
    Novelty & 
    $\uparrow$ & $\upuparrow$ & $\upuparrow$ & $\upuparrow$ & $\uparrow$ \\
    \midrule
    \textbf{Category} & 
      \multicolumn{5}{c}{\textbf{\textit{Most Liked (\%)}}} \\[1ex]
    Ease of Use                             & 20 & 27 & 27 & 43 & 17 \\
    Instructions                            & 27 & 20 & 13 & 13 & 27 \\
    Remote Voting                           & 23 & 10 & 20 & 10 & 13 \\
    Security                                & 3  & 20 & 10 & 17 & 23 \\
    QR Codes                                & 7  & 13 & 17 & 3  & 10 \\
    Other (Positive)                        & 3  & 10 & 10 & 7  & 10 \\
    Other                                   & 17 & 0  & 3  & 7  & 0  \\
    \midrule
    \textbf{Category} & 
      \multicolumn{5}{c}{\textbf{\textit{Most Disliked (\%)}}} \\[1ex]
    Process Complexity                      & 27 & 23 & 30 & 23 & 23 \\
    Credential Handling                     & 20 & 23 & 7  & 17 & 30 \\
    Confusion                               & 7  & 17 & 17 & 13 & 17 \\
    Security / Coercion                     & 7  & 10 & 23 & 13 & 7 \\
    In-Person Reg.                          & 7  & 3  & 7  & 0  & 3 \\
    (None)                                  & 17 & 13 & 10 & 23 & 10 \\
    Other                                   & 17 & 10 & 7  & 10 & 10 \\
    \midrule
    \midrule
    \textbf{Types} & 
      \multicolumn{5}{c}{\textbf{\textit{Use Errors (\# Events)}}} \\[1ex]
    Tore Receipt                & 3 & 1 & 0 & 0 & 1  \\
    Almost Tear                 & 3 & 4 & 0 & 1 & 0  \\
    Envelope Pick               & 1 & 2 & 0 & 1 & 1  \\
    Discarded Real Cred.        & 0 & 0 & 0 & 0 & 1  \\
    Activate Difficulty        & 3 & 5 & 6 & 5 & 10  \\
    \vspace{0.5em}
    \textit{Total} (48)         & 10 & 12 & 6 & 7 & 13 \\
    \midrule
    \textbf{Levels} & 
      \multicolumn{5}{c}{\textbf{\textit{Use Errors (\# Events)}}} \\[1ex]
    Mistakes                    & 10 & 7 & 5 & 7 & 11 \\
    Violations                  & 0  & 5 & 1 & 0 & 2 \\
    \midrule
    \textbf{Types} & 
      \multicolumn{5}{c}{\textbf{\textit{Use Errors (\# Required Facilitator Events)}}} \\[1ex]
    Kiosk               & 4 & 1 & 0 & 0 & 2 \\
    Activation          & 2 & 3 & 3 & 2 & 2 \\
    Study-Wide          & 0 & 0 & 0 & 0 & 1 \\
    \midrule
    \midrule
    \textbf{Reporting} & 
      \multicolumn{5}{c}{\textbf{\textit{Kiosk Reporting Rate}}} \\[1ex]
    Facilitator & 
    0\% & 0\% & 10\% & 0\% & 47\% \\
    Survey & 
    10\% & 7\% & 20\% & 7\% & 57\% \\
    \bottomrule
\end{tabularx}
\caption{\textbf{Usability Results Overview.}
The top section up to ``Like Least''
represents participants' perception of system usability
and user experience. %
The middle section represents their
use errors and facilitator interventions during the study.
The final distinct table represents the kiosk 
reporting rate.
$\upuparrow$ is top 25\% of benchmark studies, $\uparrow$ is top 50\%, %
$\downarrow$ is bottom 50\% and $\downdownarrow$ is bottom 25\%.
}

\label{tab:usability:results}
\end{table}

We present the results from participants' single-word
summaries and the system aspects they liked or disliked
the most.
For the single-word summaries,
we classified each word based on sentiment,
marking them as positive, neutral or negative.
\Cref{tab:single-word} presents the words that occur
more than once.
We also devised categories that best encapsulate
the items expressed and split them into two groups:
most liked and most disliked~(\Cref{tab:usability:results}).

We observe that the 
control group perceived the system most
positively.
As the level of engagement increased,
subsequent groups showed a consistent decline
in positive ratings.
Meanwhile,
neutral and negative ratings vary between 13 to 27\%,
with the exception of group M (a 37\% negative rating).
This surge partly derives from participants'
suspicion of the kiosk's unexpected (and incorrect) behavior,
expressing their surprise with words like ``hacked'' and ``suspect''---terms
absent from other groups' feedback, including SM.
Moreover,
while group M shows the fewest participants
valuing security among the treatment groups,
group SM shows the most.
Despite these observations,
we cannot confirm statistical differences 
between the groups and the ratings 
(Fisher's Exact Test, $p = 0.24$).

We find a roughly equal divide between those who find
the system easy to use and those who consider it complex.
We also observe a roughly equal split between 
those who appreciated the system's instructional guidance
and those who disliked the handling 
(\eg scanning, storage) of paper credentials.
We also find that
given the lack of online voting in the study location,
15\% of participants appreciate the ability to vote remotely.
A minority (4\%) expressed dissatisfaction
with the need for in-person registration.
Finally,
we observe that participants in the treatment groups
valued the system's security at least
threefold compared to the control group,
with half of these comments praising fake credentials.

\subsection{Use Errors}\label{sec:usability:use_errors}
\begin{figure*}[!t]
    \centering
    \begin{subfigure}[t]{0.45\textwidth}
        \centering
        \includegraphics[width=\linewidth]{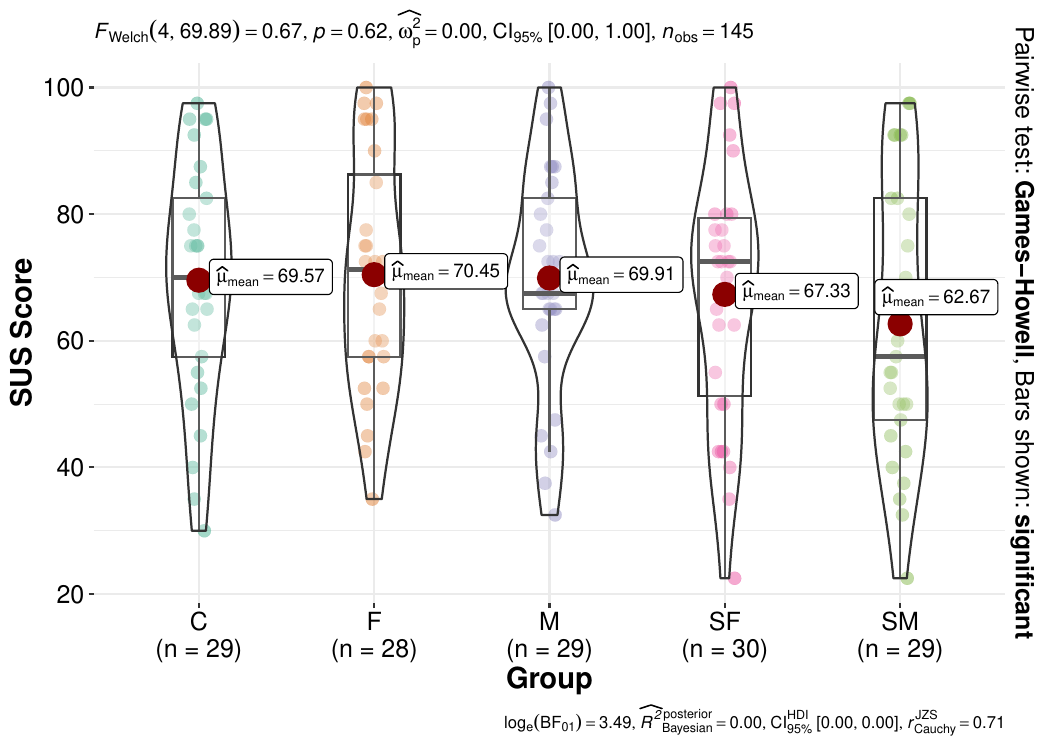}
        \caption{\textbf{System Usability Scale Across Groups}}
        \label{fig:sus}
    \end{subfigure}
    \hfill
    \begin{subfigure}[t]{0.54\textwidth}
        \centering
        \includegraphics[width=\textwidth]{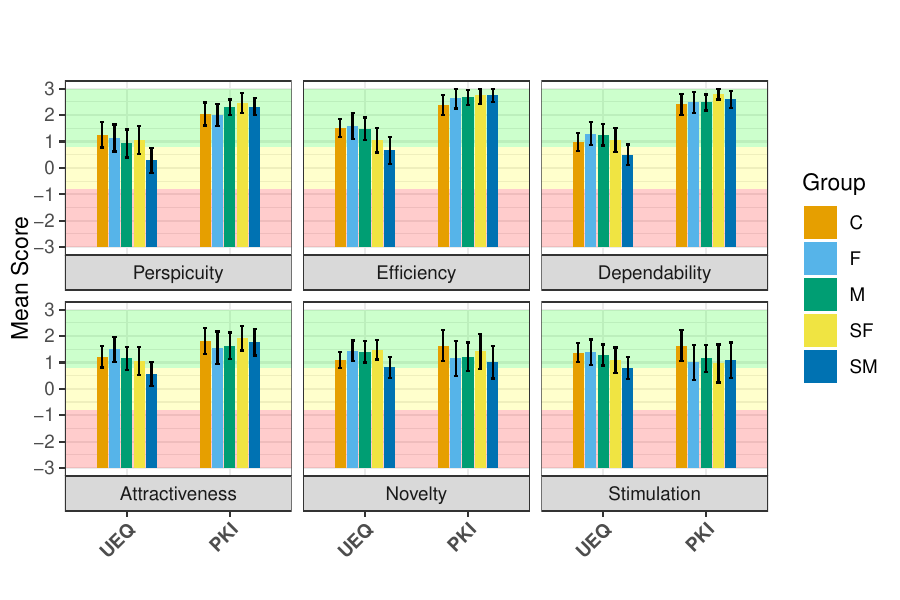}
        \caption{\textbf{User Experience Questionnaire Mean Scores Across Scales} 
        }
        \label{fig:ueq}
    \end{subfigure}
    \caption{\textbf{Usability Scales}}
    \label{fig:usability_metrics}
\end{figure*}

\noindent
During the study,
we record observational notes for each participant,
including use errors and facilitator interventions.
We focus on %
use errors fundamental to the credentialing process,
while we discuss interface and device-induced use errors 
in~\cref{apx:misc:system:other-use-errors}.
We classify use errors into two categories:
mistakes and violations.
We characterize mistakes when participants'
plan or intended action is flawed,
typically resulting from misinterpretation of instructions.
In contrast, violations arise when participants intentionally
disregard or skip instructions.
We now present a discussion of process-induced use errors we observed,
as reported in~\Cref{tab:usability:results}.

\parhead{Credentialing.}
During the real credential creation process with an honest kiosk (C, F, SF),
5 participants (6\%) prematurely tore off the receipt after
the kiosk printed the first QR code---a
mistake that required us to intervene during activation.
An additional 8 participants (9\%) attempted to tear off the receipt
but when they encountered resistance from the receipt printer, 
they rectified their mistake
by consulting the kiosk display guiding them to pick
and scan an envelope.
In terms of envelope selection,
5 participants (6\%) initially opted for an envelope
that did not correspond with the symbol on the receipt;
the kiosk alerted them to this mistake and they all
successfully scanned a correct envelope on their second attempt.
A single participant misinterpreted the 
``discard check-in ticket'' screen after creating their real credential
and mistakenly threw away their credential.

\parhead{Activation.}
The majority of use errors occurred during the activation phase,
after ``leaving the government office'',
with approximately 19\% of participants encountering difficulties
and 8\% requiring facilitator assistance.
Participants commonly skipped the on-screen instructions, a violation error.
Participants therefore removed the entire receipt from the envelope
when the device prompted to scan three QR codes.
These participants likely associated the ``three QR codes'' to be
the three QR codes on the receipt rather than the two QR codes from
the receipt and one QR code from the envelope.
For those participants who read the instructions yet still faced challenges,
mistakes often involved scanning only a subset of QR codes at once instead
of all three simultaneously, and placing the credential on the device's
screen instead of on the table to be scanned by the device's camera.
Considering these activation errors, 
it naturally follows that credential handling emerged as the second 
most disliked aspect of the system among participants.

\parhead{Success Rate.}
95\% of participants succeeded in registering
and creating their credentials without assistance from the facilitator.
This result appears to support the practicality and usability
of a coercion-resistant voter-verifiable registration process like TRIP's,
despite its complexity.
Success rate drops when we include errors later in the voting pipeline, however.
92\% of participants activated their credential without help.
In combination, 87\% accomplished both registration and activation by themselves,
with 19 participants (13\%) needing assistance.
Counting participants who mistakenly used
their fake credential to cast their vote
further reduces the success rate to 83\%,
as we discuss in~\cref{sec:credentials:distinguish}.

\parhead{Statistical Analysis.}
We categorized participants into four age groups: 
18-30, 31-45, 46-65, and 65+. 
Analysis of use errors across these age groups using the chi-squared test 
revealed no statistically significant differences ($p = 0.8791$). 
Further examination employing logistic regression to assess the 
interaction effects between age groups, ethnicity, education, and gender 
on use errors also indicated no statistically significant associations 
as determined by the Wald test.

\parhead{System Improvements.}
The study results suggest several potential ways
to improve the success rate.
A receipt detection mechanism could enable the kiosk
to restart credential creation
if the user prematurely tears off the receipt.
Activation issues might be reduced
by reading the QR codes incrementally,
instead of expecting three readable QR codes in one image.
Redisplaying the activation instructions
upon unsuccessful activation may help
users who skip the instructions.
An animation showing how to use
the device's rear-facing camera to scan credentials
might also help.

\parhead{Study Comparisons.}
We compare our success rate with
three usability studies containing 
seven variants of voting systems:
the only other JCJ-style coercion-resistant 
voting study we are aware of by Neto et al.~\cite{
neto2018CredentialsDistributionUsability},
a study by Acemyan et al.~\cite{
acemyan2014UsabilityVoterVerifiable}
on three voter-verifiable systems
(Helios~\cite{adida2008Helios}, 
Prêt à Voter~\cite{ryan2009PrEtVoterVoterVerifiable},
Scantegrity II~\cite{chaum2009ScantegrityIIEndtoEnd}),
and a recent study on STAR-Vote~\cite{
acemyan2022STARVoteUsability},
a secure, auditable and transparent ballot marking device 
for in-person voting.
The Neto et al.~\cite{
neto2018CredentialsDistributionUsability} between-subjects
study involved 80 university-affiliated individuals aged
18-39 acquiring their real credential using one of the following
three variants and then casting both real and fake votes:
(1) in-person acquiring a pen drive, 
(2) remotely via email, and 
(3) in-person with a password set by the voter. 
The success rate for the first two variants is 100\% 
and the last one is 85\%.
Acemyan et al.~\cite{
acemyan2014UsabilityVoterVerifiable}
conducted a within-subjects study with 37 diverse participants
on Helios, Prêt à Voter and Scantegrity II where Helios is
an voter-verifiable online voting system while Prêt à Voter 
and Scantegrity II are coercion-resistant, 
voter-verifiable, in-person voting systems.
The success rate 
for Helios is 60\%,
Prêt à Voter is 60\%, and
Scantegrity II is 50\%.
The study on STAR-Vote~\cite{
acemyan2022STARVoteUsability} reported
a success rate of 93\%.

\parhead{Discussion.}
TRIP outperforms three out of the seven variants,
including prominent coercion-resistant, voter-verifiable, in-person
voting systems Prêt à Voter and Scantegrity II,
which TRIP most closely resembles as a coercion-resistant,
voter-verifiable, in-person registration system.
Despite the remaining four variants surpassing TRIP,
they either fall short in achieving one or more of these properties---%
illustrating the challenge in designing a 
usable coercion-resistant system---%
or have a homogeneous study population.
TRIP substantially improves this success rate to 83\%,
up from 60\% (Prêt à Voter) and 50\% (Scantegrity II),
and closer to the 93\% success rate that STAR-Vote achieves.

To our knowledge, NIST has established standards for
electronic voting~\cite{NIST2021VVSG}, 
but no official success rate exists.
Only a NIST internal document~\cite{NIST2007UsabilityBenchmarks} 
recommends a success rate of 98\% (known as Total Completion Score) 
but where 98\% only needs to fall within the 95\% confidence interval.

\subsection{Perceived Usability and User Experience}
\label{sec:usability:results}

We now present the system usability scale and
the user experience questionnaire scores and 
compare them to benchmarks, along with
the systems mentioned in~\cref{sec:usability:use_errors}.

\parhead{System Usability Scale.}
In \Cref{fig:sus},
we present the System Usability Scale (SUS) 
scores across our study groups,
while removing 5 participants
due to inconsistent answers
(agreeing to most positive and negative items),
as suggested by Sauro~\cite{SUSGuide}.
We observe that groups C, F, and M demonstrate
similar average and median scores while groups
SF and SM exhibit lower average scores,
with group SM having a score decrease of 10\%
over the control group.
Despite these observations,
Welch's one-way ANOVA reveals no statistically 
significant differences (p = 0.62).
Nonetheless,
we can compare our SUS mean scores with a benchmark consisting
of 446 studies involving a range of products and services,
from business/consumer software to hardware devices~\cite{SUSGuide}.
These 446 studies reveal an average SUS score of 68,
with standard deviation of 12.5.
Groups C, F, and M achieve
marginally superior SUS scores, ranking in the
55.1st, 57.8th and 56.1st percentiles, respectively.
Conversely, groups SF and SM underperform with scores
ranking in the 47.8th and 35.0th percentiles.
Works such as Bangor et al.~\cite{bangor2009SUSRating} 
have proposed adjective ratings based on the SUS score.
These ratings would classify group F---representing
the expected common case
in realistic settings---as
``Acceptable,'' and achieving a ``Good'' adjective rating.

\parhead{SUS: Study Comparisons.}
We now compare group F's SUS score with
Neto et al. and Acemyan et al.'s study~\cite{
acemyan2014UsabilityVoterVerifiable,
neto2018CredentialsDistributionUsability}.
While Neto et al.'s variant 2 and 3 achieve a higher mean SUS score
of 77.5 (81st percentile) and 77.4 (81st percentile), respectively,
these were not statistically significantly higher than group
F's score (Welch one-sided t-Test, $p = 0.094$ and $p = 0.051$).
Neto et al., unfortunately, did not administer the
SUS questionnaire to participants for variant 1,
which is the most comparable to TRIP,
in that it involves
the in-person delivery of voter materials.
From Acemyan et al.'s~\cite{
acemyan2014UsabilityVoterVerifiable} study\footnote{
These numbers are not present in the text~\cite{
acemyan2014UsabilityVoterVerifiable}; we
extract these numbers from their figure 4 and assume
a 95\% confidence interval.},
Group F's mean SUS score of 70.4 was statistically 
significantly higher than Prêt à Voter's score of 61 
(Welch one-sided t-test, $p < 0.05$), and %
Scantegrity II's score of 59 
(Welch one-sided t-test, $p < 0.05$) %
but lacked conclusive statistical difference 
from Helios' score of 76
(Welch two-sided t-test, $p = 0.17$).

\parhead{SUS: Discussion.}
The comparable percentile scores and 
the absence of statistical difference 
between groups C, F, and M suggest 
that introducing fake credentials do not
change participants' perceptions of usability. 
Further, 
even with voters engaging in a nontrivial 4-step protocol, 
F achieved an Acceptable rating and 
achieved statistically significantly higher
SUS scores than Scantegrity II and Prét à Voter.

\parhead{User Experience Questionnaire.}
\Cref{fig:ueq} and \Cref{tab:ueq:scores} illustrate
the UEQ scores for each scale
(Attractiveness, Dependability, Efficiency, Novelty,
Perspicuity, and Stimulation),
along with the UEQ Key Performance Indicator (KPI) scores,
which represent participants' perception of an
ideal registration experience.
Traditional usability aspects encompass 
Efficiency, Perspicuity, and Dependability, 
while Novelty and Stimulation relate to user experience.
Upon comparison, 
participants view this system as
slightly surpassing their expectations 
in Novelty, Stimulation, and Attractiveness.
However, it falls short in the traditional usability aspects
by approximately a full point.
Nonetheless, scores above -2 and 2 are extremely rare 
due to differing opinions and answer tendencies~\cite{
schrepp2023UEQHandbookV10}
(\eg avoidance of extreme responses).
Typically, as depicted in~\Cref{fig:ueq},
a positive evaluation is a score above 0.8,
neutral evaluation is a score between -0.8 and 0.8,
and negative evaluation is a score less than -0.8~\cite{
schrepp2023UEQHandbookV10}.
Based on this metric,
Group F has a positive evaluation for each of the scales.

\parhead{UEQ: Benchmark.}
Similar to SUS,
to better assess usability,
we need to compare the UEQ scores with other studies.
We first compare our scores to the UEQ benchmark~\cite{
schrepp2023UEQHandbookV10},
consisting of 21,175 participants from 468 diverse studies.
While group F's scores for efficiency, stimulation and novelty
rank in the top quartile, the perspicuity
scores are between the 50th and 75th percentiles.
This suggests that despite participants finding it
more challenging to familiarize themselves
with this system compared to the benchmark,
they still
complete their tasks without excess effort,
while perceiving the system as both engaging and innovative.

\parhead{UEQ: Study Comparisons.}
We now compare our scores with 
other voting-related studies,
although the number of studies that administer
UEQ is more limited.
One comparable study conducted by Marky et al.~\cite{
marky2019UsabilityCodeVoting}
focuses on voter verifiability---though not coercion-resistance---%
through the use of code voting~\cite{
ryan2013PrettyGoodDemocracy}
as prominently used in the Swiss Online Voting System~\cite{
2021SwissPostProofs}.
This approach provides voters with physical materials,
typically by mail, to cast a vote so as to prevent
potential vote alteration by the voter's device.
Marky et al. tests three voting code variants---%
manual codes, QR codes, and tangibles---with 18
participants.
Our interest lies in the QR-codes variant,
as it is most similar to TRIP, and is also
the option favored in Marky et al.'s study.
Unlike in TRIP, 
their scores for user experience (stimulation and novelty) are neutral,
although their scores for the traditional usability aspects 
(perspicuity, efficiency and dependability) are all positive.
Marky et al.'s study also achieves higher perspicuity and efficiency
scores, 2.2 vs. 1.13, and 2 vs. 1.59, respectively, while TRIP
achieves a marginally higher dependability score 1.3 vs. 1.2.

\subsection{Detecting a Malicious Kiosk}
\noindent
We present our findings from participants
exposed to our malicious kiosk (groups M and SM). 
We assess their ability to identify the kiosk's 
misbehavior during their interaction with the kiosk
and via a survey question probing if they
detected any irregularities while creating their 
real credential~(Appendix \ref{apx:survey:credentials}).
We administered the same survey question to groups F and SF
as a control to identify false positives rates.
\Cref{tab:usability:results} presents the reporting rates.

\parhead{Kiosk Reporting Rate.}
Among group M participants, 10\% reported the kiosk's misbehavior
to the facilitator, whereas this rate significantly increased to 47\%
for group SM.
Further, this difference in reporting rate is statistically significant 
(Chi-squared, $\chi^2 = 8.2079, p < 0.01$; 
Cramér's V $= 0.4068$).\footnote{For Cramér's V,
values around 0.1 are typically considered small, 
around 0.3 indicate a moderate effect size, and 
values of 0.5 or greater signify a strong effect.}
Instructional video 3, shown to group SM,
thus substantially increased the reporting rate.

\parhead{Survey Reporting Rate.}
A greater percentage of participants reported the kiosk's misbehavior
in the exit survey: 20\% for group M and 57\% for group SM.
We still observe a statistically significant difference in the reporting
rates between groups M and SM (Chi-squared test, $\chi^2 = 7.0505, p < 0.01$;
Cramér's V $= 0.3771$).
Unlike the kiosk reporting rate,
however, the survey reporting rate was non-zero for groups C, F, and SF,
averaging at 8\%; these participants primarily cited confusion about the process.

\parhead{Real-World Scenario.}
According to participants' responses to our survey 
questions,
85\% are willing to watch an instructional video before 
participating in voter registration in a real world context.
However, only 59\% indicated willingness to ``lock up'' their
personal devices in a locker before entering the booth 
and retrieve them afterwards.

\parhead{Study Comparison.}
Bernhard et al.~\cite{
bernhard2020VotersDetectManipulation}
conducted a study among diverse participants to
examining the rate at which voters could detect
malicious ballot manipulation from ballot marking
devices (BMDs).
Both our and Bernhard et al.'s studies involved
participants operating under the election authority's
supervision, interacting directly with an official device:
the BMD in their case and the kiosk in ours.
In both studies, we assess whether participants can
visually discern anomalous behavior from these devices.
In their study, without any guidance,
only 6.6\% of 31 participants
reported the error to the facilitator.
When participants were asked before submission
if they had carefully reviewed their ballot,
12.9\% of 31 reported the error to the facilitator.
In contrast,
the reporting rate for TRIP is 10\% without security priming,
and 47\% with priming.

\section{Usability of Fake Credentials}\label{sec:coercion}

\noindent
This section evaluates the comprehension and usability of
fake credentials among the 120 participants exposed to them.

\subsection{Understanding Fake Credentials}
\newcolumntype{C}[1]{>{\centering\arraybackslash}m{#1}}
\begin{table}[]
    \centering
    \footnotesize
    \begin{tabularx}{\linewidth}{
    >{\centering\arraybackslash\hsize=3.38\hsize}X
    >{\raggedleft\arraybackslash\hsize=0.16\hsize}X
    >{\raggedleft\arraybackslash\hsize=0.16\hsize}X
    >{\raggedleft\arraybackslash\hsize=0.16\hsize}X
    >{\raggedleft\arraybackslash\hsize=1.25\hsize}X
    >{\raggedleft\arraybackslash\hsize=1.39\hsize}X
    >{\raggedleft\arraybackslash\hsize=0.5\hsize}X}
        \toprule
            Quiz Attempts                    &  1  & 2  & 3 & Continue 
            (Missing Correct Options) & Continue (Incorrect Option) & Count \\
            \midrule
            Real Credential Steps   &  65 & 21 &  1 &   0  &  3  & 90 \\ %
            Real Credential Usage  & 148 &  2 &  0 &  0   &  0  & 150 \\
            Test Credential Usage            & 84  & 18 &  4 &  7   &  7  & 120 \\ %
            Distinguish Credentials          & 91  &  1 &  0 &  0   & 0  & 92 \\
        \bottomrule
    \end{tabularx}
    \caption{\textbf{Quiz Results.}
    This table represents the quiz attempts;
    ``Continue'' signifies that these participants could not 
    pass the quiz after the third attempt.
    ``Missing Correct Options'' indicates that these
    participants only selected correct answers but did not
    simultaneously choose all correct answers.
    ``Incorrect Option'' indicates that these participants
    selected the incorrect option in one of the quiz attempts
    and could not rectify it by the third attempt.
    \cref{sec:methods:quizzes} discusses participants' assignment
    to quizzes.
    }
    \label{tab:quiz-results}
\end{table}

\noindent
We examine voter comprehension of fake credentials
in two distinct stages:
first with a pop quiz presented by the kiosk,
then later in the survey.
This dual-stage assessment
measures understanding of passive instruction
and efficacy of active instruction.
The quiz poses the question
``Which of the following purposes can you use a test credential for?''
(\cref{apx:study:quizzes}).
Not selecting ``To cast a vote that counts in election''
underscores the crucial understanding that fake credentials cannot
be used to cast a real vote; the remaining three correct options
must be concurrently selected to pass the quiz.

\parhead{Quiz Results: Test Credential Usage.}
We find that 70\% of participants answer the quiz
correctly on their first attempt~(\cref{tab:quiz-results}).
An additional 15\% succeed on the second attempt, including
3 who initially chose the incorrect answer.
The third and final attempt sees the remaining 15\% (18 participants),
split as follows: 4 participants answer correctly,
7 fail to select all correct options simultaneously,
and 7 choose the incorrect option.
In this quiz stage,
92\% of participants
avoid selecting the incorrect answer in all attempts,
thereby demonstrating an understanding of fake credentials.

\parhead{Exit Survey.}
Among the 8\% of participants (3 from second attempt + 7 from third attempt)
who incorrectly selected
``To cast a vote that counts in an election'' in the quiz,
five of them correctly identified the use of
fake credentials in the survey.
In the remaining five,
four could not recall the purpose of fake credentials, and
one only wrote ``voting''.
We infer that 4\% (5 participants) likely finished
the study without a clear understanding of fake credentials,
with one participant thinking that fake
credentials can be used to cast real votes.
Among the participants who did not select the incorrect option,
89\% remembered, in their free-text responses,
the use of fake credentials to resist coercion.
The remaining 11\% wrote about their use to educate others,
test the system, and
even profit by selling their fake credentials.

\subsection{Usability of Real and Fake Credentials}
\label{sec:credentials:distinguish}
\noindent
Considering that both real and fake credentials are used
to cast votes in the same way---%
assessed previously in~\cref{sec:usability:use_errors}---%
the success of using real or fake credentials
ultimately hinges on the voter's ability to distinguish between the two.
In the study, out of the 120 participants exposed to fake credentials,
92 chose to create one or more fake credentials.

\parhead{Credential Distinguish Quiz.}
We initially verify participants' understanding of how
to distinguish their credentials by administering the
Credential Distinguish quiz~(\cref{apx:study:quizzes}).
Remarkably, 99\% of participants select the correct option
``Only myself with my pen markings'' on their first attempt,
thereby confirming their apparent comprehension~(\cref{tab:quiz-results}).

\parhead{Distinguishing Credentials.}
We assess participants' \emph{actual} ability to distinguish their
credentials by asking them to cast a vote using
their real credential after completing voter registration.
Among the 92 participants, 90\% (83)
accurately identified their real credential.
For the 10\% (9 participants) who failed
to do so, we explore the reasons behind this.

One participant discarded their real credential during registration,
as mentioned in~\cref{sec:usability:use_errors}.
We examined each set of credentials the remaining eight participants created
to ascertain whether their real credentials had distinct markings,
and found that they did.
For details about how
study participants generally marked their credentials,
see Appendix \ref{apx:misc:fake:markings}.
Although credential marking was apparently not the cause of these errors,
many other potential causes remain that we could not identify,
such as memory lapses, misreading instructions,
environmental distractions,
or even deliberately disobeying our instructions
(\eg viewing the facilitator as a potential coercer,
which ``by our own game'' might suggest
voting with a fake credential in the facilitator's presence).\footnote{
We informally observed a few participants hiding their credentials 
under the table out of the facilitator's sight, 
which might suggest such a "facilitator as adversary" perspective.}

To gain deeper insights into individuals' ability
to recall sensitive data, we consider password-related studies~\cite{
brown2004GeneratingRememberingPasswords,
dhamija2000DejaVu},
which indicate a wide range of retention rates from 23\% to 98\%.
With a 90\% success rate, this rate lies in the upper end
of this spectrum and is similar to Déjà Vu~\cite{dhamija2000DejaVu},
a study conducted on using images for authentication.
Voters marking their own credentials incorporate
several known retention-enhancing strategies, such as
user-generated content and
favoring visual imagery over text.
The Déjà Vu study provides encouraging evidence that
imagery-based memory degradation is significantly lower than that of PIN/passwords:
after one week, only 1 out of 20 participants failed to login with images while
7 to 6 individuals failed to login with PIN and passwords, respectively.

We expect that the duration
over which TRIP users must remember their credential markings
should normally be much shorter than the requirements for long-term passwords:
from leaving the privacy booth until credential activation on their device.
This duration may be only minutes
if the voter brought their voting device with them,
hours or at most days for a credential the voter activates at home or elsewhere.
Voters who do forget which credential is real
may re-register at any time to obtain fresh credentials.

\parhead{Confidence in Distinguishing Credentials.}
Participants also rated how confident
they were in recalling their real credential on
a 7 point scale, where 1 is ``No confidence''
and 7 is ``Extremely  confident.''
87\% of participants gave a confidence rating of 5 or more,
with 55\% of participants giving the extremely confident
rating.
7\% rated a 4 (neutral), and ratings 1 and 2 each got one participant.

\parhead{Fake Credentials In Reality?}
We ask participants about their willingness
to create fake credentials if such a system existed;
53\% of participants affirm that they would.
Testimonials vary from ``because I have been in
situations where others forced me to vote'',
``to argue less with people voting for
another candidate'' to ``[...] a fun souvenir'',
and ``I'm in a demographic
where I cannot imagine having someone
trying to solicit my vote, if they did,
I would simply tell them no without fear.''
Additional testimonials, including a taxonomy, are available
in Appendix~\ref{apx:misc:fake:create}.

\parhead{Discussion.}
Despite introducing an unforeseen and unprecedented concept
to 120 participants, only 4\% (5 participants) appear to have finished
the study without a reasonable grasp of the use of fake credentials.
Furthermore,
although voters were not required to generate fake credentials,
76\% of participant chose to create at least one.
This proactive engagement is further expressed by 53\% of participants
who are willing to create fake credentials in a real-world context.
Lastly,
in spite of the identical nature of fake and real credentials,
90\% of participants successfully distinguished between the two
when aiming to activate and cast a real vote,
on par with a study of using
images for authentication.

\section{Discussion}\label{sec:discussion}

\noindent
This section primarily offers key takeaways of our 
findings, building on the many results we have 
presented earlier.
We also expand on our study's limitations.

\smallskip
\parhead{The Need For Coercion-Resistance.}
Our study reveals that a quarter of participants have either 
personally experienced coercion, or is aware of someone who has.
These findings
highlight the necessity of coercion-resistance to 
uphold free and fair elections.
In essence,
votes should not just be private but
incorruptible.
This importance is further reflected with these same 
participants rating~(\cref{fig:voting-methods:coercion-happened}) 
statistically significantly lower trust levels 
for both mail-in and online voting methods compared
to in-person voting (with BMD).

\smallskip
\parhead{Usability of Coercion-Resistant Voting Systems.}
Designing a usable coercion-resistant voting 
system is a difficult task.
As seen in~\cref{sec:usability:use_errors},
voting systems without coercion resistance, 
such as STAR-Vote~\cite{bell2013STARVote},
exhibit much higher success rates than systems that address coercion,
such as Prêt à Voter~\cite{khader2013PretAVoterReceiptFree} 
and Scantegrity II~\cite{chaum2009ScantegrityIIEndtoEnd}
(93\% versus 60\% and 50\% respectively).
This disparity has motivated other coercion-resistant 
strategies with less intricate tasks 
such as deniable re-voting~\cite{
achenbach2015JCJDeniableRevoting,lueks2020VoteAgain},
where voters only receive a single voting credential
and perform the same vote casting process to override 
their previous vote.
However, this strategy is vulnerable to last-minute coercion,
as well as to domestic coercion---%
the primary source of coercion in our findings---%
where the coercer simply keeps the voting credential or device.
The 83\% success rate achieved by TRIP shows promise
in narrowing the success gap between coercion-resistant and
non-coercion-resistant systems,
though this gap remains an important usability challenge for future work.

\smallskip
\parhead{The Impact of Security Vigilance on User Experience.}
Our findings show an interesting tradeoff 
between user experience and security vigilance.
Participants from group SF typically rated system usability 
and user experience lower than group F, 
despite the only change being an 
educational brief in the instructional video about
the rare possibility of a compromised 
kiosk~(\cref{apx:study:instructional-videos}).
This increase in discomfort, however,
came with a statistically significant improvement in 
detecting a malicious kiosk in
group SM compared with group M (47\% vs. 10\%),
with no increase in false positives.
Meanwhile, twice as many participants in SM
compared to M liked the system's security 
the most.
These observations confirm that finding the right balance
between security education and user comfort
is an important challenge.

\smallskip
\parhead{Limitations.}\label{sec:limitations}
\noindent
This study has several important limitations.
First,
it did not evaluate long-term questions,
such as whether voters can effectively
store, manage, and properly use credentials on their devices after activation
to vote in successive elections over periods of years.
While the problem of remembering which paper credential is real
technically ends at credential activation time in TRIP,
the problem of remembering which \emph{device} holds the real voting credential --
or of remembering which account, password, or PIN guards it -- remains.
Thus, the detailed design and evaluation of long-term storage systems
for coercion resistance with fake credentials
remains an important area for future work.
Systematically evaluating the usability of long-term credential storage systems
may be a particularly difficult challenge,
as studies addressing this challenge
would necessarily be longitudinal in nature,
requiring participants to remain involved over an extended period.
We know of no deployed voting system that could facilitate such a study.

The TRIP system's accessibility to voters with 
disabilities is also important but beyond the scope of this study.
In the future, a thorough design for accessibility
and corresponding study should include 
voters with disabilities, which would likely yield
further insights. %
Although we aim for diversity,
the study being conducted in a single geographical
location may affect the generalizability of our
findings.
Finally,
the study did not replicate
a government office setting, 
which might influence differences
in participants' views and behaviors.
Similarly, the study does not replicate
circumstances where a voter is truly under coercion.

\section{Related Work}\label{sec:related}
\noindent
In previous sections (\cref{sec:usability:use_errors} 
\& \cref{sec:usability:results}),
we compared our results with other usability studies.
This section therefore provides a broader overview
of related work.

A few works~\cite{neumann2012CivitasRealWorld,
neumann2013SmartCards2,estaji2020UsableCRSmartCards}
have considered usability issues with
fake credentials and proposed 
possible implementations,
such as using smart cards.
These works did not conduct
systematic studies with real users, however.

Civitas~\cite{clarkson2008Civitas},
strengthens the verifiability of the JCJ coercion-resistance scheme
by having voters
interact with multiple registration officials. 
There are no user studies of Civitas, however,
and such a study may be difficult
due to the logistical challenges presented by multiple registrars,
and the absence of a well-defined voter-facing design.

Other works have looked into using other
countermeasures to resist coercion~\cite{kulyk2020HumanFactorsCoercion}, 
such as deniable re-voting and masking.
The masking approach~\cite{ryan2013PrettyGoodDemocracy}
provides each voter with a unique value $b$,
known only to them, which is then used to
offset their cast vote.
However, a usability concern arises as it remains
uncertain if voters can recall and apply their assigned
value $b$ during vote casting~\cite{kulyk2020HumanFactorsCoercion}.

An important security-related study in voting is one by
Distler et al.~\cite{distler2019EVotingSecurityVisibility},
which examines the impact of displayed security mechanisms on
user experience.
They found that participants exposed to these mechanisms, \eg
messages like ``Encrypting your vote'', have a
better user experience than those who were not exposed
to these mechanisms.
Their research thus highlights the importance
of the visibility of security measures.
In contrast to their work, we study the involvement
of voters to achieve security measures.

\section{Conclusion}\label{sec:conclusion}

This paper presented a study with 150 individuals 
to evaluate their experiences and perceptions of coercion, 
along with their views of a coercion-resistant system
with a voter-verifiable registration system resilient to 
coercion via the use of fake credentials.
A quarter of participants 
had either personally faced coercion 
or know of someone who did. 
Remarkably, 96\% of participants understand the concept 
of fake credentials, with 90\% successfully 
casting a vote using their real credential. 
Moreover, 
over half of the participants exposed to fake credentials
indicated their willingness to use them in reality.
These findings show promise in narrowing the usability gap
between voting systems with and without coercion-resistance, 
and confirm the need for continued research into 
making verifiable coercion-resistant voting systems more usable.

\subsection*{Acknowledgements}
The authors would like to thank the study participants
for their invaluable insights and their time, as well
as the reviewers for their helpful feedback.
This project was supported in part by 
the armasuisse Science and Technology, the AXA Research Fund, 
and the US ONR grant N000141912361.

\newpagetoggle

\ifdefined\ieee
\bibliographystyle{support/IEEEtran}
\else
\bibliographystyle{plain}
\fi
\bibliography{
bib/lhm-zotero-without-url.bib,
bib/custom.bib,
bib/dedis/comp.bib,
bib/dedis/fault.bib,
bib/dedis/lang.bib,
bib/dedis/net.bib,
bib/dedis/os.bib,
bib/dedis/priv.bib,
bib/dedis/sec.bib,
bib/dedis/soc.bib,
bib/dedis/theory.bib}

\appendices

\section{Miscellaneous Questions}
\label{apx:misc}
\begin{figure*}[!t]
    \centering
    \begin{subfigure}[t]{0.47\linewidth}
        \includegraphics[width=\linewidth]{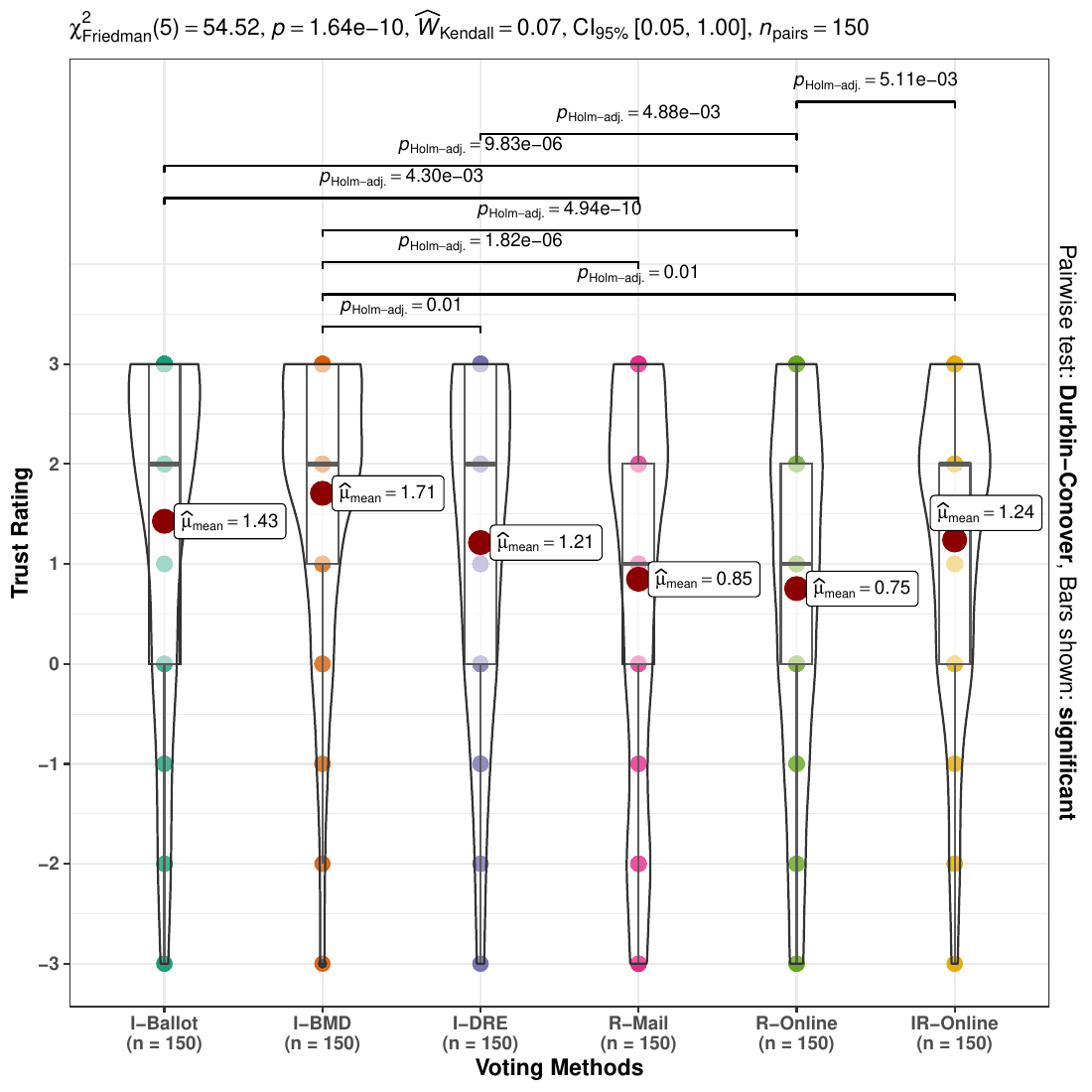}
        \caption{\textbf{Voting Methods Trust Rating from All Participants.}
        All 150 participants' trust ratings across voting methods.
        }
        \label{fig:voting-methods:all}
    \end{subfigure}
    \hfill
    \begin{subfigure}[t]{0.47\linewidth}
        \includegraphics[width=\linewidth]{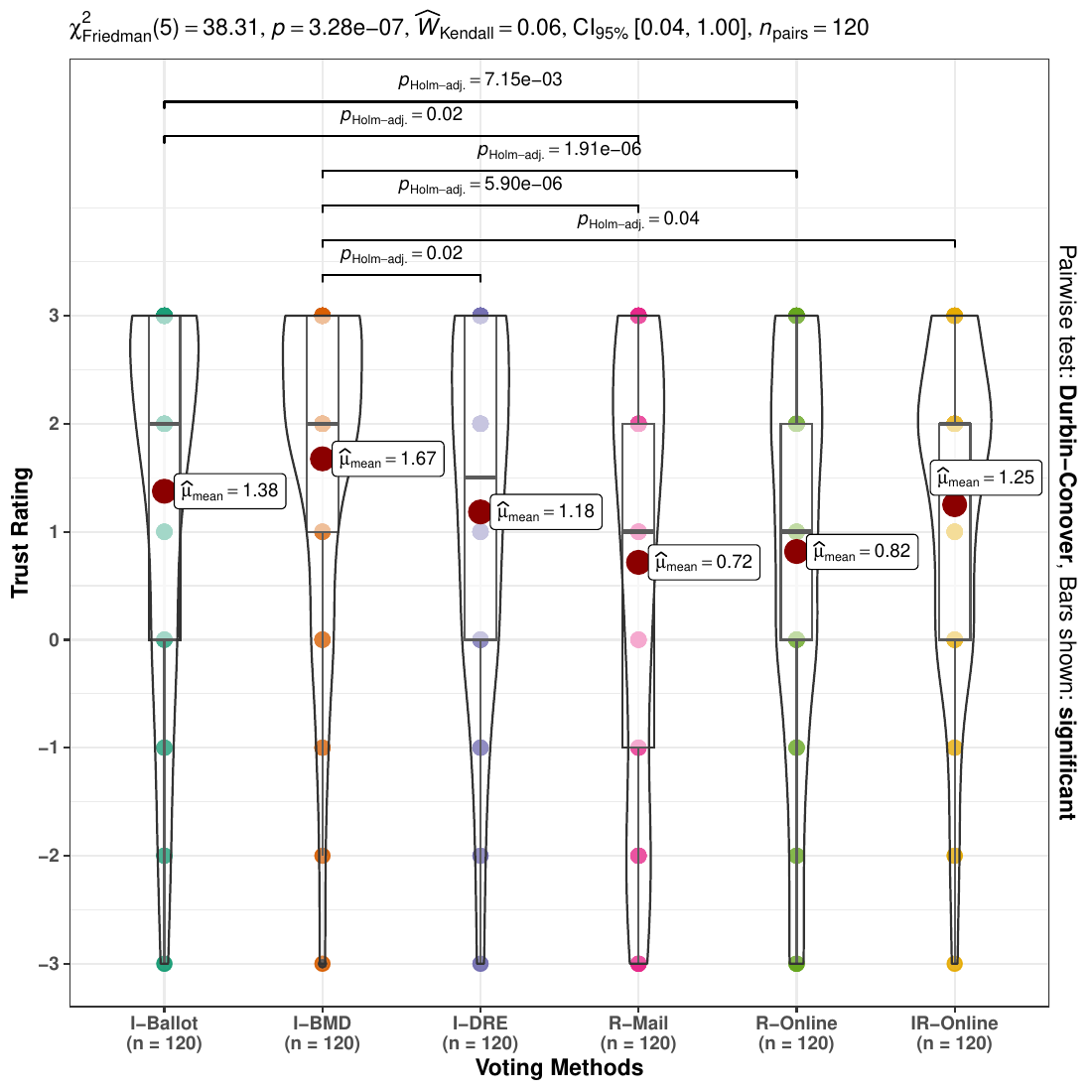}
        \caption{\textbf{Voting Methods Trust Rating from
        Participants Exposed to Fake Credentials.}
        The trust ratings for various voting methods, as given
        by the 120 participants exposed to fake credentials.
        }
        \label{fig:voting-methods:fake-credentials}
    \end{subfigure}
    \hfill
    \begin{subfigure}[t]{0.47\linewidth}
        \includegraphics[width=\linewidth]{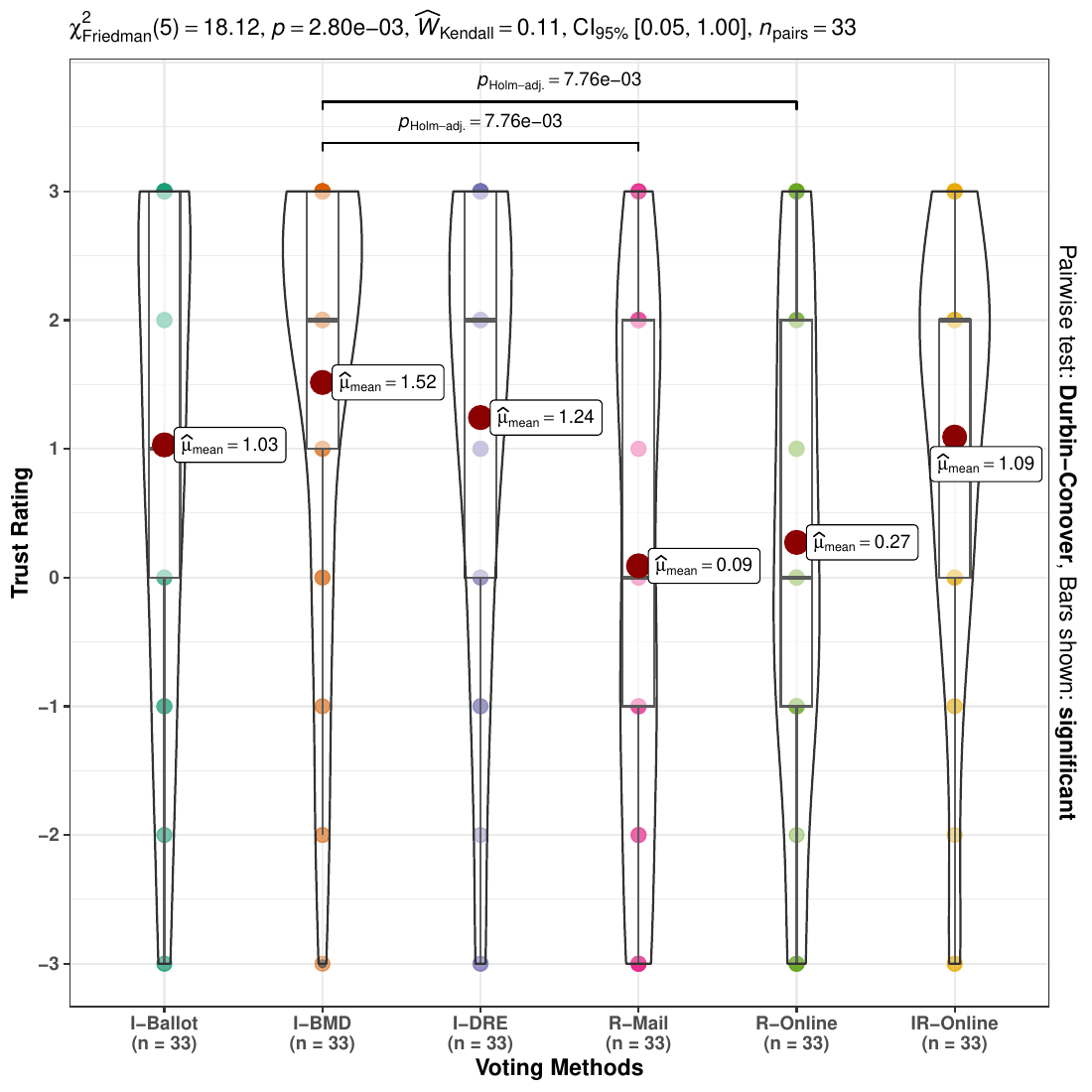}
        \caption{\textbf{Voting Methods from Coercion-Happened Participants.}
        The trust ratings for various voting methods, as given
        by the 39 participants who experienced coercion or know someone who
        has experienced coercion.
        }
        \label{fig:voting-methods:coercion-happened}
    \end{subfigure}
    \hfill
    \begin{subfigure}[t]{0.47\linewidth}
        \includegraphics[width=\linewidth]{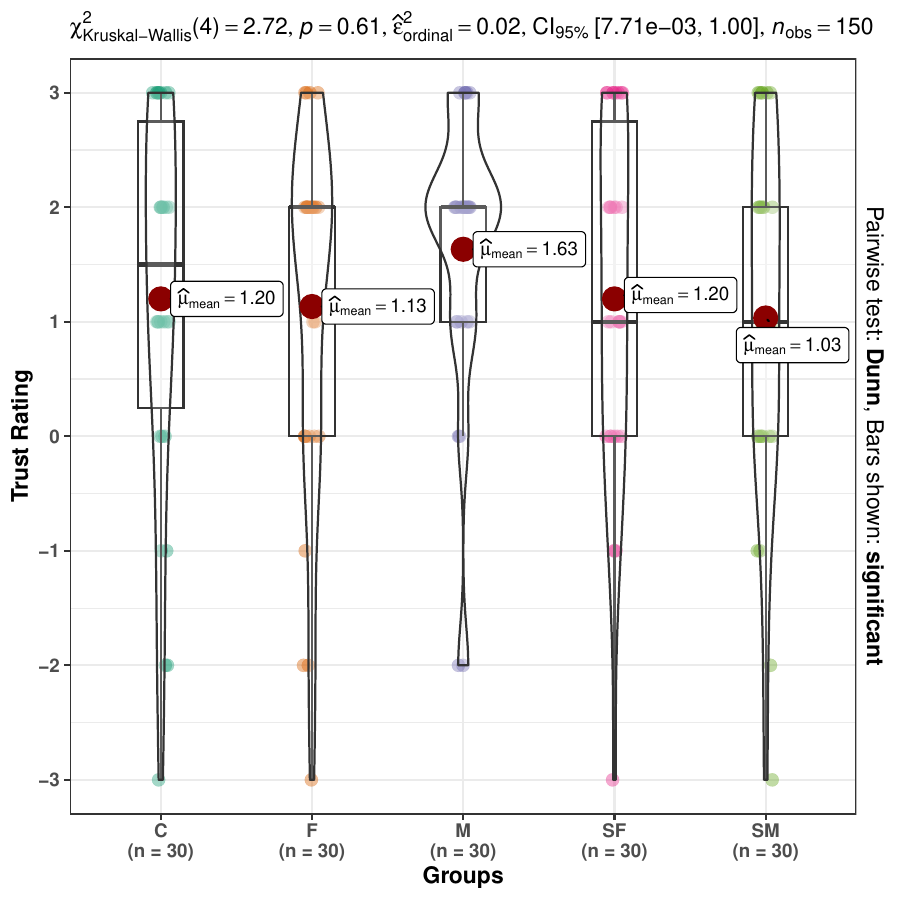}
        \caption{\textbf{IR-Online Voting Method Across Groups.}
        The participants' trust ratings across groups for the online voting with
        in-person registration voting method.
        }
        \label{fig:voting-methods:ir-online}
    \end{subfigure}
    \caption{\textbf{Voting Methods.}}
    \label{fig:voting-methods}
\end{figure*}

\subsection{Trust Rating in Other Voting Methods}\label{apx:misc:voting-methods}
\noindent

\parhead{Most Trusted.}
Participants express the most trust towards \texttt{I-BMD},
with 69\% deeming it as highly trustworthy or
higher, and assigning it an average trust
score of 1.68 out of 3 (\cref{tab:voting-methods}).
Moreover,
\texttt{I-BMD} is statistically significantly higher than
that of all other voting methods,
except for \texttt{I-Ballot}~(\cref{fig:voting-methods:fake-credentials}), which
earns the same level of trust from 60\% of participants and
an average trust score of 1.38.

\parhead{Least Trusted.}
Participants express the least trust in \texttt{R-Mail},
only 46\% considering it
highly trustworthy or better (mean score of 0.72).
Participants also view \texttt{R-Online} similarly,
yielding similar mistrust (mean score of 0.85).

\parhead{All Participants.}
\Cref{fig:voting-methods:all} depicts the trust ratings across voting methods from all 150 participants (including the control group),
along with our statistical results.
Participants trust in-person registration 
for online voting statistically significantly more than they trust 
a fully-online voting system.
Participants further emphasized this during the study by asking us
whether we would require individuals to verify their eligibility, 
as we had not required this in the study.

\parhead{Discussion.}
46\% of voters
cast an absentee ballot in the 2020
U.S. presidential election~\cite{
atske2020VotingMethods2020Presidential}.
Despite this,
mail-in voting has the second-lowest
trust score (\cref{fig:voting-methods:all})
or the lowest trust score
(\cref{fig:voting-methods:fake-credentials} \& \cref{fig:voting-methods:coercion-happened}).
Participants expressed a lack of
confidence in the U.S. postal service's capacity
to deliver mail reliably.
Despite concerns with BMDs~\cite{
stark2019SecurityConcerns,appel2020BallotMarkingDevices},
including a study finding that only 6.6\% of participants
can detect a \emph{manipulated} printed ballot~\cite{
bernhard2020VotersDetectManipulation}, \texttt{I-BMD}
received the highest trust score across voting methods.

\subsection{Use Errors}\label{apx:misc:system:other-use-errors}

We discuss process-induced errors in
\cref{sec:usability:use_errors}.
This section discusses
device- or interface-induced errors, which
include premature scanning of check-in tickets,
unawareness to answer the quiz,
and unfamiliarity with the device's camera.

\parhead{Premature Scanning.}
Of the 150 participants,
56 scanned their check-in ticket before
the kiosk asked them to do so.
This occurred 51 times in groups C,
F and SF\@.
47 (84\%) corrected their mistake without our
intervention upon recognizing the correct intended action,
such as pressing ``Continue'' or ``Begin''.
No participants in groups M or SM required our intervention.
This pattern occurred mostly in groups C, F and SF due
to an instructional slide, for which groups M and SM
were not privy due to the kiosk's malicious setting.
Upon viewing this instructional slide's four steps,
starting with ``Scan Check-in ticket'',
participants scanned their ticket instead of pressing ``Begin''
to initiate the process.
When we were asked to intervene,
we guided participants towards the intended action,
typically by pressing ``Begin'', although 84\% of
participants did not require help.
These errors may result
from a mix of environmental factors,
such as inadequate contrast between
the button and the screen in an outside environment,
alongside behavioral factors.
Such behavioral factors may include participants
scanning their check-in ticket before
reading the instructions,
perhaps due to over-confidence
after seeing the instructional video.
To address these issues,
we propose permitting a combination of inputs:
\eg
interpreting the scanning of the check-in ticket
during an instructional slide as an indication
that the participant is prepared for subsequent steps.

\parhead{Mobile Device-induced.}
An additional 12 participants had initial difficulty
with activating their credential as they either placed
the QR codes directly on the device's screen or used the front-facing camera.
In each instance, we intervened to instruct the participant to use
the back-facing camera.
Participants who continued to need assistance beyond
this guidance are reported in the main body of the paper.

\parhead{Kiosk-Induced.}
Three participants faced technical difficulties
with the kiosk device, necessitating our intervention.
Two of these instances arose from participants not
realizing they had to complete a quiz before proceeding,
while one participant had trouble locating the QR code scanner.

\subsection{Credential Markings}\label{apx:misc:fake:markings}
\begin{table}[t]
    \centering
    \footnotesize
    \begin{tabularx}{\linewidth}{
    >{\centering\arraybackslash\hsize=1.96\hsize}X
    >{\raggedleft\arraybackslash\hsize=1.38\hsize}X
    >{\raggedleft\arraybackslash\hsize=0.84\hsize}X
    >{\raggedleft\arraybackslash\hsize=0.91\hsize}X
    >{\raggedleft\arraybackslash\hsize=0.81\hsize}X
    >{\raggedleft\arraybackslash\hsize=0.6\hsize}X
    >{\raggedleft\arraybackslash\hsize=0.9\hsize}X
    >{\raggedleft\arraybackslash\hsize=0.6\hsize}X}
        \toprule
        Differentiator & Envelope Symbol & Number & Scribble & Symbol & Text & Type Change & Count \\
        \midrule
        Explicit       &        0        &   0    &   0      &    0   &   30 &  2   & 32 \\
        Implicit       &        0        &   0    &   0      &    5   &   4  &  10  & 19 \\
        Indistin-guishable  &        5        &   1    &   10     &    8   &   9  &  4   & 37 \\
        \midrule
        Count          &        5        &   1    &   10     &    13  &   43 &  16  & 88 \\
        \bottomrule
    \end{tabularx}
    \caption{\textbf{Credential Markings.} Type of markings
    that participants used to differentiate their fake credentials from their real
    credential. ``Type Change'' signifies participants use distinct types, ``Scribble''
    includes signatures, and ``Envelope
    Symbol'' indicates participants used no markings, instead relying on
    distinct envelope symbols. Four sets of envelopes are unaccounted for.}
    \label{tab:credentials}
\end{table}

We examine how participants mark their credentials to
distinguish their real credentials from their fake ones.
We categorized the set of credentials
for 88 participants (four sets were unaccounted for),
first by looking at the type of marking and then 
level of differentiation based on patterns across credentials.
We find six general types of markings:
\begin{itemize}
    \item Envelope Symbol: No pen markings.
    \item Number: Random numbers.
    \item Scribble: Indecipherable writings, including signatures.
    \item Symbols: Predominantly shapes such as stars or squares but sometimes
        including smiley faces or animal drawings.
    \item Text: Words like ``Real'' or ``Test'' or participants' initials.
    \item Type Change: Alternation between two or more of the above categories
        across the real and test credentials.
\end{itemize}

Our findings, detailed in \Cref{tab:credentials},
show that 49\% of participants used \textit{Text}
to differentiate their credentials, followed by
\textit{Type Change} at 18\%.
Furthermore,
42\% of participants marked their credentials indistinguishably,
making it impossible
to differentiate the real credential from the fake ones.
Participants may lie about their real
credential by marking their real one as ``Fake'' and their
fake one as ``Real'' but participants in our study were
not influenced to do so.

\subsection{Fake Credentials in Reality?}\label{apx:misc:fake:create}
We examine whether participants exposed to fake credentials
would be willing to create them alongside their
real credential if such a system existed in the real world.
53\% would do so,
citing the following reasons:
\begin{itemize}
    \item Security (27 participants):
        ``Because I have been in situations where others forced me to vote'',
        ``On the off chance someone tries to force me to vote in a particular
        way''.
    \item Convenience (15 participants):
        ``no harm in creating a test credential'',
        ``might come in handy'', and
        ``to argue less with people voting for another candidate''.
    \item Education (7 participants):
        ``It might be useful to show my students what the process was like'',
        ``To see that it worked and educate family and friends'', and
        ``It would be good for teaching people how to vote.
        Also might be a fun souvenir''.
    \item Other (12 participants):
        ``I would want to create as many test credentials as possible before the
        market became flooded with test credentials'', and
        ``I’d like to test my device setup''.
    \item No Reason Given (2 participants)
\end{itemize}

\noindent
The following were reasons for not creating a test credential:
\begin{itemize}
    \item Unnecessary (39 participants):
    ``I'm in a demographic where I cannot imagine having someone trying to solicit my vote,
    if they did, I would simply tell them no without fear,'' and
    ``Not interested in the uses.''
    \item Cumbersome or Confusing (11 participants):
    ``I don't want to take the risk of being confused
    between my real credentials and the fake one.
    I can use videos or websites to teach someone else how to vote,'' and
    ``More to lose, and I would mix them up.''
    \item No Reason Given (7 participants).
\end{itemize}

\section{Meta-Review}

The following meta-review was prepared by the program committee for the 2024
IEEE Symposium on Security and Privacy (S\&P) as part of the review process as
detailed in the call for papers.

\subsection{Summary}
This paper presents a field-experiment and usability test of the TRIP (Trust-limiting In-Person Registration) protocol: a voter-verifiable registration system for coercion-resistant online voting via the idea of fake credentials. The core contribution of the paper is to provide a sense of how usable the TRIP system is with an experiment that attempts to capture as wide of a voter-pool as possible. The paper details the protocol, discusses the recruitment and study strategy, and provides a detailed analysis of how users interacted with the system, what confusion points and errors occurred, and how these results might be incorporated into a TRIP-enabled system in the future.

\subsection{Scientific Contributions}
\begin{itemize}
\item Independent Confirmation of Important Results with Limited Prior Research
\item Provides a Valuable Step Forward in an Established Field
\item Provides a New Data Set For Public Use
\item Creates a New Tool to Enable Future Science
\item Establishes a New Research Direction
\end{itemize}

\subsection{Reasons for Acceptance}
\begin{enumerate}
\item The paper presents a valuable step forward in an established field by presenting results from a field-experiment for a potential online voting scheme. The authors do a good job of detailing the user study and rigorously testing the TRIP mechanism.
\item The paper also offers researchers insight into leveraging fake credentials for coercion resistance and outlines potential pain points of this mechanism in real-world deployments
\end{enumerate}

\subsection{Noteworthy Concerns} %
\begin{enumerate} %
\item The results illustrate that some voters may encounter difficulties using fake credentials, with 17\% of users encountering an issue and 10\% of users mistakenly voting with their fake credential. These findings could adversely affect the system’s deployability and underline the need for further research to mitigate and reduce such use errors.
\item It is unclear how requiring in-person registration may impact voter turnout. This presents a practical consideration for the real-world deployment of this system that must be examined in future research.
\item The system's formal threat model excludes coercion threats that could be realistic in practice, such as side-channel attacks, electronic surveillance via wearable devices a voter is successfully coerced to sneak into the registration booth, or a coercer waiting to strip-search the voter immediately upon leaving the booth to confiscate all of their voting credentials. The authors suggest a potential countermeasure for this last case, but all of these risks beyond the formal threat model – and potential mitigations for them – remain important areas for further study.
\end{enumerate}

\section{Study Materials}
This section provides details
about the study design.

\parhead{Additional notation.}
Considering the differences between groups,
we adopt specific notations to indicate the
text each group encounters.
The notation  $[X|Y]$ indicates that $X$
is the content present for the control group
while $Y$ is the content present for the treatment groups.
On the other hand, $[X\$Y]$ indicates that $Y$ is the content present for the malicious kiosk groups (M, SM)
while $X$ is the content present for the other groups.

\subsection{Facilitator's Scripts}\label{apx:study:materials:scripts}

\noindent

\parhead{Recruitment.}
Would you like to participate in a usability study on online voting?
The study takes approximately 30 minutes and you will be compensated
\$20 for your time.
To be eligible, you must have registered to vote in the past.
The whole process is fictitious, you would be voting on your
favorite color or favorite sport.

\parhead{Welcome.}
Thank you for your interest in our study.
Online voting is also known as electronic voting
or e-voting and enables voters to cast votes
using their own device in the comfort of their own home.
Here is the information sheet.
We also have an instructional video for you to watch.
At any time during or after the study,
you may withdraw from participation by
contacting us using the contact info on the information sheet.
Please review the information sheet and
sign the consent form if you agree to participate in the study.
I will keep the original signed consent form.
You may take a copy of the consent form or take a
picture for your records. The information sheet is yours to keep.

\parhead{Groups C \& F \& M Scenario Priming.}
Imagine that the year is 2030.
Online voting is now commonplace.
You are renewing your drivers license at a government office.
You see a poster announcing that you can vote online.
You are interested, so you watch an instructional video on how
to register for online voting.
After watching the video, you go to the clerk to get started.

\parhead{Groups SF \& SM Scenario Priming.}
Imagine that the year is 2030.
Online voting is now commonplace.
You are renewing your drivers license at a government office.
You see a poster announcing that you can vote online.
You are interested,
so you watch an instructional video on how to register for online voting.
You worry that hackers are trying to manipulate U.S. elections,
but the video teaches you how to register securely and
protect your vote from manipulation.
After watching the video, you go to the clerk to get started.
Being security conscious, you monitor the registration process carefully
and report anything suspicious.

\parhead{Check-In.}
Here is your check-in ticket,
you may now enter the booth.

\parhead{Check-Out.}
Please hand me [\textbar any one of] your credential[\textbar s]
to perform the check-out process.

\parhead{Activation.}
Here is your credential.
Now imagine that you have left the government office and are at home.
Please use this device as the device that you trust for online voting.
For the purposes of this study,
activate your real credential and
then cast a vote in our mock elections.
[\textbar You may then activate your test credentials if you wish]

\parhead{Conclusion.}
That's it.
In practice, you would be able to vote in all elections
for the next 5-10 years on your device until you need to
renew your identification documents,
where you would repeat this signup process.

\subsection{Instructional Video Slides}
\label{apx:study:instructional-videos}
\begin{figure}[t]
    \centering
    \begin{subfigure}[b]{\linewidth}
         \centering
         \includegraphics[width=\textwidth]{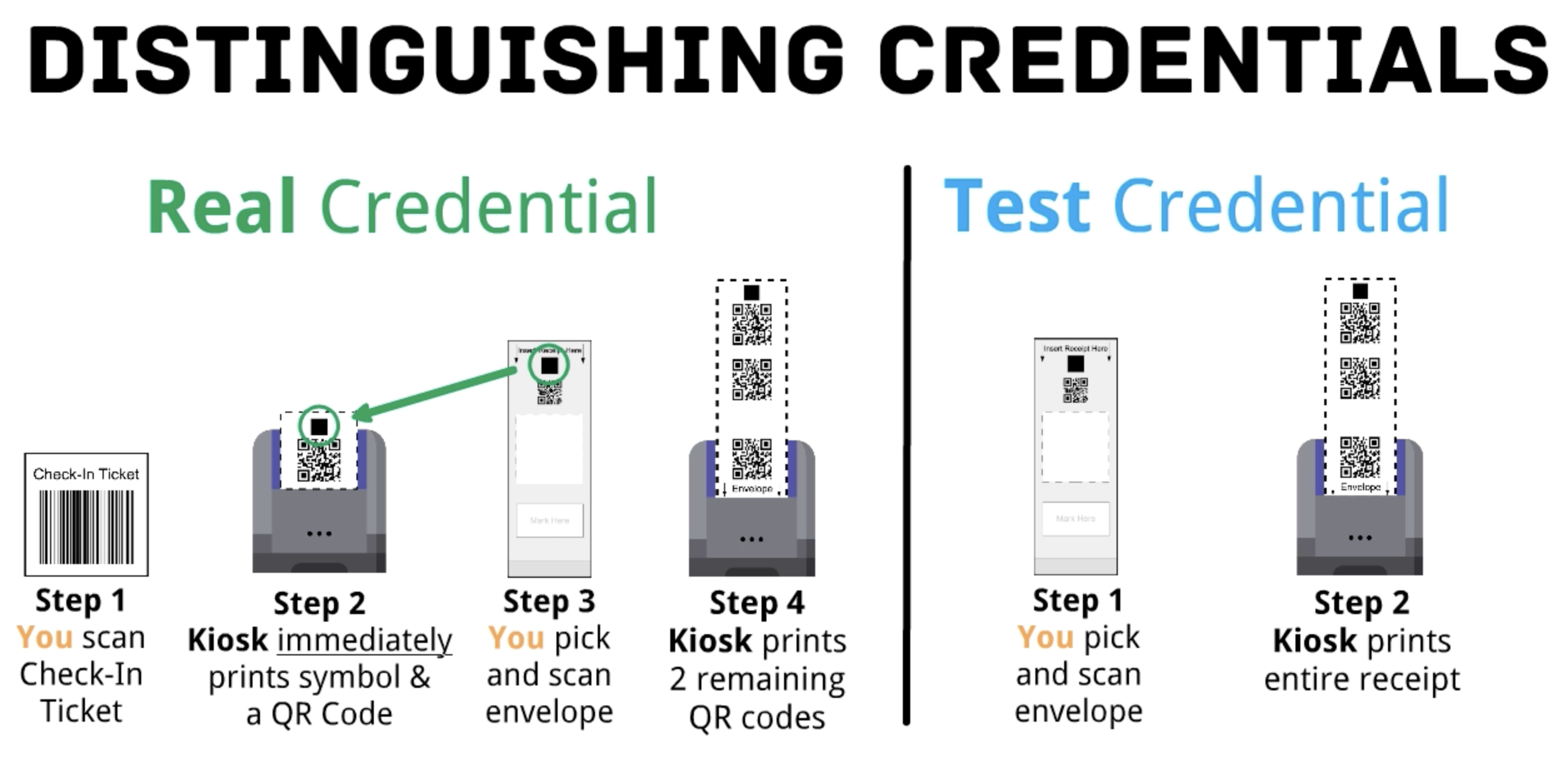}
         \caption{\textbf{Video 2.}
         Participants are exposed to a slide that shows,
         side by side, the creation of a real credential and of a fake credential.}
         \label{fig:video:b:creds}
    \end{subfigure}
    \hfill
    \begin{subfigure}[b]{\linewidth}
         \centering
         \includegraphics[width=\textwidth]{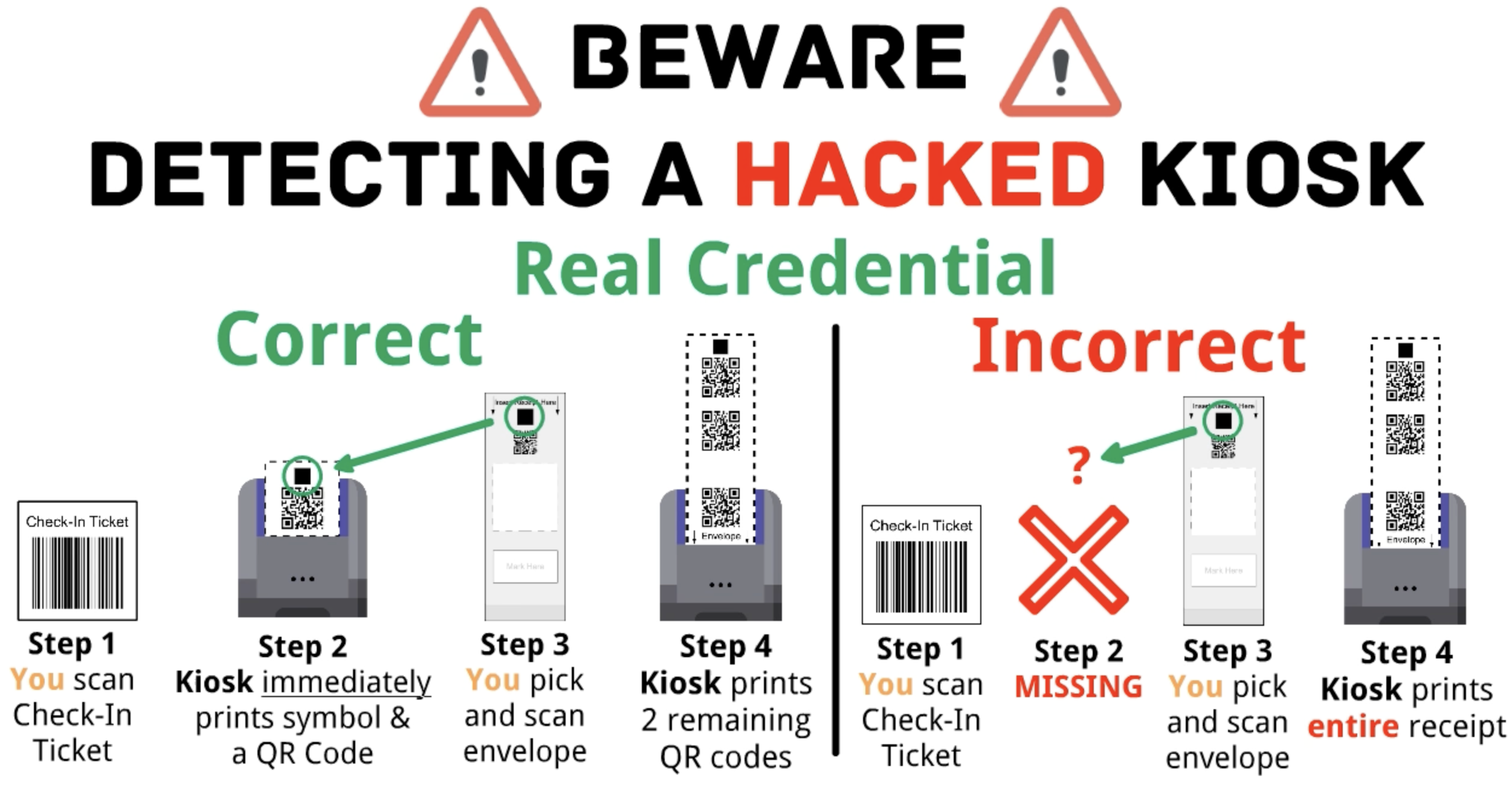}
         \caption{\textbf{Video 3.}
         Participants are exposed to a slide that
         which depicts how a kiosk can trick the voter
         into creating a fake credential instead of a real
         credential.}
         \label{fig:video:c:creds}
    \end{subfigure}
    \caption{\textbf{Instructional Video Differences.}
        Screenshots from the instructional video
        concerning how to distinguish real and fake credentials.}
    \label{fig:video:distinguish-credentials}
\end{figure}

\Cref{fig:video:distinguish-credentials} presents
the distinguish credential slides between
instructional videos 2 and 3, as introduced in~\cref{sec:study:videos}.

\subsection{Quizzes}
\label{apx:study:quizzes}
We present our quizzes and italicize
the correct option(s).
In the study,
the kiosk shuffles the options
for each participant.

\parhead{Real Credential Steps.}
While creating your [\textbar real~]voting credential,
when do you pick and scan an envelope to the kiosk?
(Check all that apply)
\begin{itemize}
    \item After a symbol and three QR codes are printed
    \item \textit{After a symbol and one QR code is printed}
    \item Before anything is printed
\end{itemize}

\parhead{Real Credential Storage.}
What should you do with your [\textbar real~] voting credential
after leaving the government office? (Check all that apply)
\begin{itemize}
    \item \textit{Activate it yourself on a device that you trust.}
    \item  Give it to someone else to activate on my behalf.
\end{itemize}

\parhead{Test Credential Usage.}
Which of the following purposes can
you use a test credential for? (check all that apply)
\begin{itemize}
    \item \textit{To teach kids about voting.}
    \item \textit{To resist pressure from someone to vote a particular way.}
    \item \textit{To sell to someone who offers to buy your real credential.}
    \item To cast a vote that counts in an election.
\end{itemize}

\parhead{Credential Distinguish.}
Who can distinguish your test credential from your real credential?
(Check all that apply)
\begin{itemize}
    \item Anyone with a device.
    \item \textit{Only myself with my pen markings.}
\end{itemize}

\section{Survey and Results}\label{apx:survey}
This section presents the exit survey
with the precise wording used,
along with any figures or tables that
could not be incorporated into the main paper.
For readability, we omit certain information
such as the response type (\eg numeric, free form,
multiple choice) and merge followup
questions to the original question.

\subsection{Background}
\label{apx:survey:background}
\begin{figure*}[t]
    \centering
    \begin{subfigure}[t]{0.58\textwidth}
        \centering
         \includegraphics[width=\linewidth]{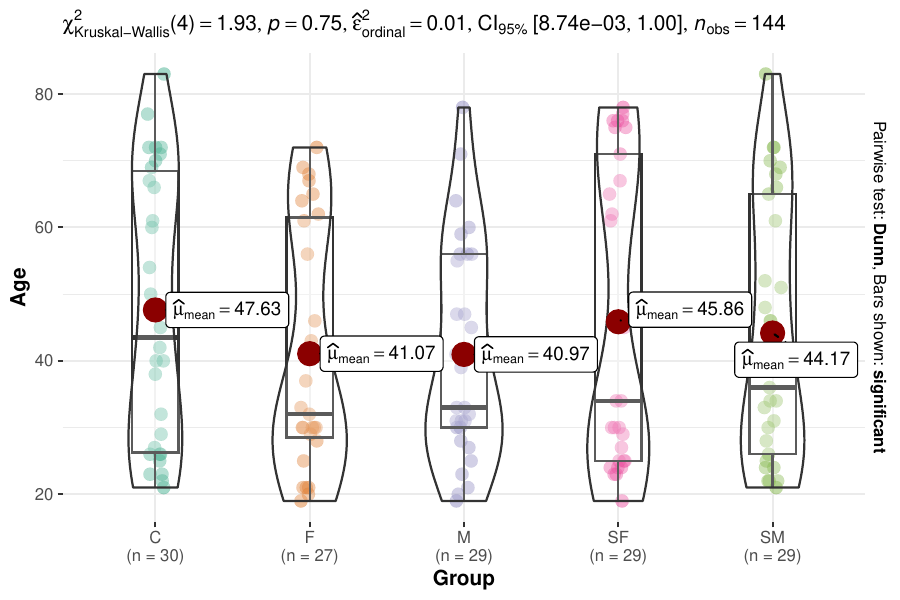}
         \caption{\textbf{Age Across Groups}.
            The age distribution of participants across groups}
         \label{fig:age}
     \end{subfigure}
     \hfill
     \begin{subfigure}[t]{0.4\textwidth}
        \centering
         \includegraphics[width=\linewidth]{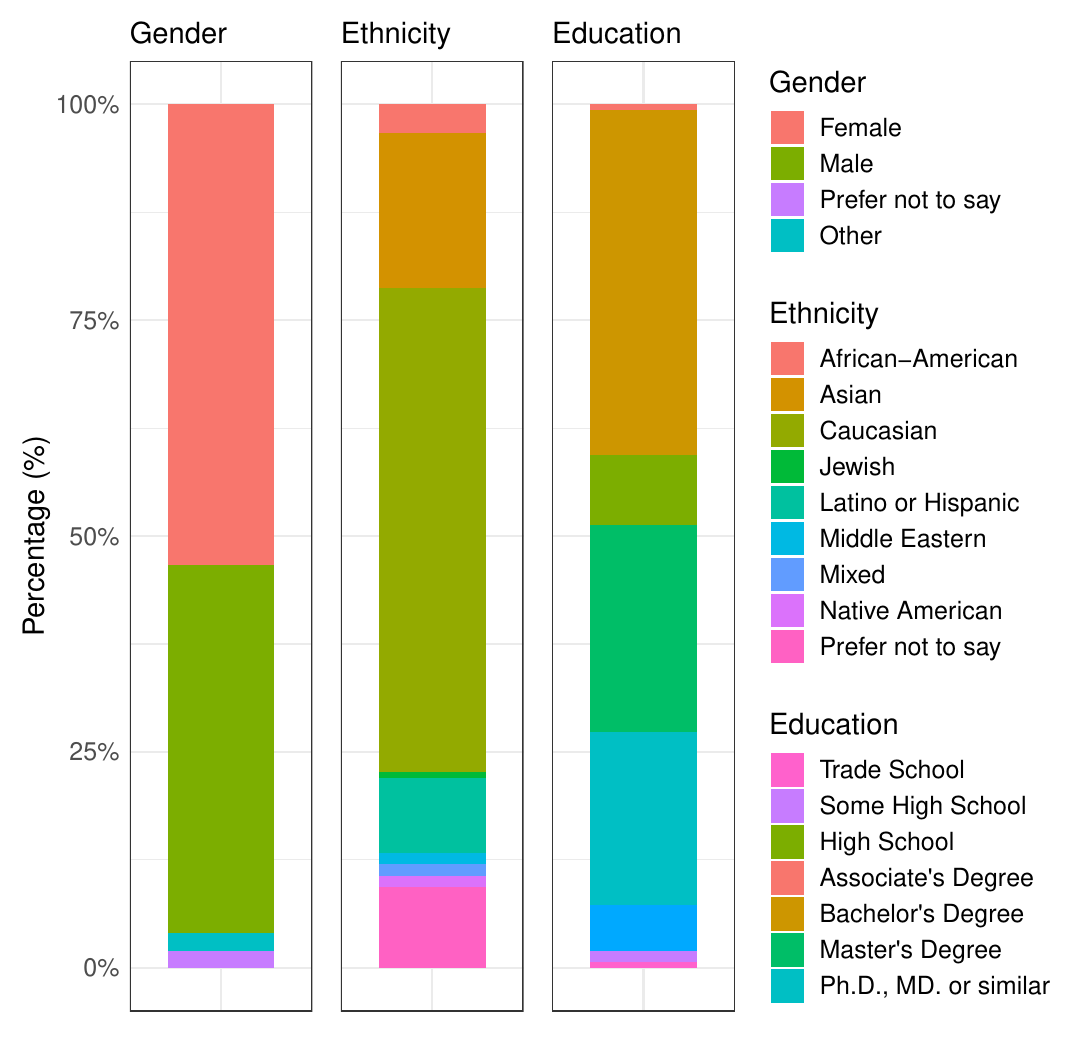}
         \caption{\textbf{Gender, Ethnicity and Education.} 
         The distribution of participants, 
         categorized by gender, ethnicity and education.}
         \label{fig:categorical-demographics}
     \end{subfigure}
     \caption{\textbf{Participant Demographics}}
     \label{fig:demographics}
\end{figure*}

\begin{enumerate}
    \item What is your age?
    \item What is your gender?
    \item What is your ethnicity?
    \item What is the highest degree or level of education you have completed?
\end{enumerate}

\noindent
\textit{\Cref{fig:age} presents the age distribution of participants
for each group and~\Cref{fig:categorical-demographics}
presents ethnicity, education and gender
distributions.}

\subsection{System Usability Qualitative}
\label{apx:survey:usability}

\begin{enumerate}
    \item Please describe your online voting signup experience
    in a single word.
    \item What elements about this voter registration system
    did you like most?
    \item What elements about this voter registration system
    did you like least?
\end{enumerate}

\subsection{System Usability Scale (SUS)}
\label{apx:survey:sus}

Please rate each statement on a scale ranging from
1 ``Strongly Disagree'' to 5 ``Strongly Agree'':

\begin{enumerate}
    \item I would like to use this voter registration
        system whenever I renew my identifications documents
        (\ie every 5-10 years)
    \item I found the system unnecessarily complex.
    \item I thought the system was easy to use.
    \item I think that I would need the support of a technical person to be able to use this system.
    \item I found the various functions in this system were well integrated.
    \item I thought there was too much inconsistency in this system.
    \item I would imagine that most people would learn to use this system very quickly.
    \item I found the system very cumbersome to use.
    \item I felt very confident using the system.
    \item I needed to learn a lot of things before I could get going with this system.
\end{enumerate}

\subsection{User Experience Questionnaire (UEQ)}
\label{apx:survey:ueq}

This questionnaire consists of pairs of contrasting attributes.
The circles between the attributes represent gradations between
the opposites. You can express your agreement with the attributes
by ticking the circle that most closely reflects your impression.

Please decide spontaneously.
Don't think too long about your decision to make sure that
you convey your original impression.

\begin{table}[!t]
    \centering
    \scriptsize
    \begin{tabular}{c c c c}
        \toprule
        Entry & Left & Right & Scale (not shown) \\
        \hline
        1 & annoying & enjoyable & Attractiveness \\
        2 & not understandable & understandable & Perspicuity \\
        3 & creative & dull & Novelty \\
        4 & easy to learn & difficult to learn & Perspicuity\\
        5 & valuable & inferior & Stimulation \\
        6 & boring & exciting & Stimulation \\
        7 & not interesting & interesting & Stimulation \\
        8 & unpredictable & predictable & Dependability \\
        9 & fast & slow & Efficiency \\
        10 & inventive & conventional & Novelty \\
        11 & obstructive & supportive & Dependability \\
        12 & good & bad & Attractiveness \\
        13 & complicated & easy & Perspicuity \\
        14 & unlikable & pleasing & Attractiveness \\
        15 & usual & leading edge & Novelty \\
        16 & unpleasant & pleasant & Attractiveness \\
        17 & secure & not secure & Dependability \\
        18 & motivating & demotivating & Stimulation\\
        19 & meets expectations & does not expectations & Dependability \\
        20 & inefficient & efficient & Efficiency \\
        21 & clear & confusing & Perspicuity \\
        22 & impractical & practical & Efficiency \\
        23 & organized & cluttered & Efficiency \\
        24 & attractive & unattractive & Attractiveness \\
        25 & friendly & unfriendly & Attractiveness \\
        26 & conservative & innovative & Novelty \\
        \bottomrule
    \end{tabular}
    \caption{\textbf{UEQ Questionnaire}}
    \label{tab:ueq}
\end{table}
\begin{figure*}[!t]
    \centering
    \includegraphics[width=0.9\linewidth]{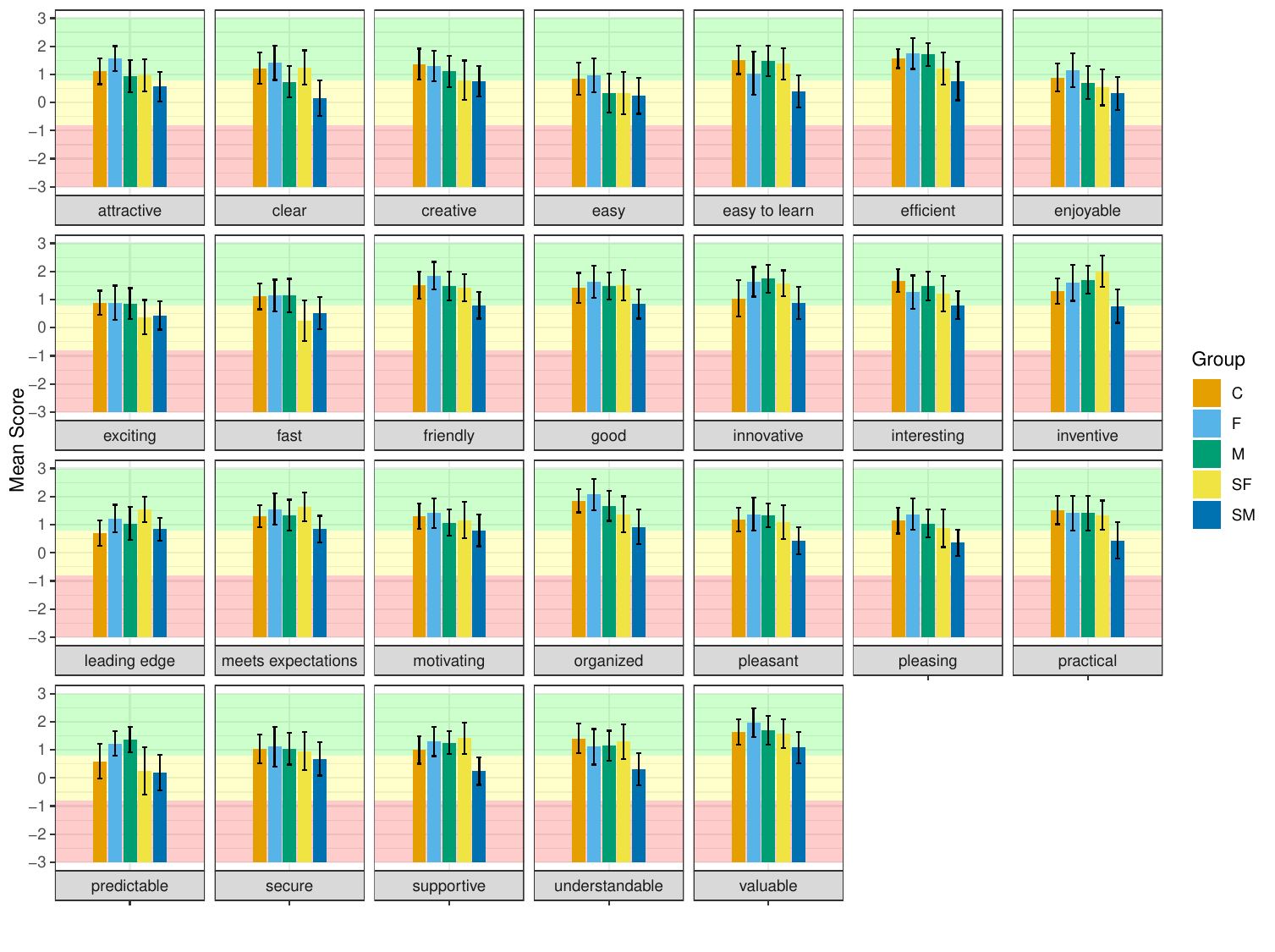}
    \caption{\textbf{User Experience Questionnaire Itemized Scores.}
    }
    \label{fig:ueq:itemized}
\end{figure*}

\begin{table*}[!t]
\small
\begin{tabularx}{\linewidth}{
    >{\centering\arraybackslash\hsize=1\hsize}X
    >{\centering\arraybackslash\hsize=1\hsize}X
    >{\raggedleft\arraybackslash\hsize=1\hsize}X
    >{\raggedleft\arraybackslash\hsize=1\hsize}X
    >{\raggedleft\arraybackslash\hsize=1\hsize}X
    >{\raggedleft\arraybackslash\hsize=1\hsize}X
    >{\raggedleft\arraybackslash\hsize=1\hsize}X
    >{\raggedleft\arraybackslash\hsize=1\hsize}X
    >{\raggedleft\arraybackslash\hsize=1\hsize}X}
    \toprule
    Group & Type & Sample Size & Attractiveness & 
        Perspicuity & Efficiency & Dependability & Stimulation & Novelty \\
    \midrule
    C & UEQ     & 27 & 1.21 (1.13) & 1.25 (1.59) & 1.51 (0.77) & 0.98 (0.85) & 1.37 (0.87) & 1.10 (0.64) \\
    C & PKI     & 27 & 1.81 (1.70) & 2.04 (1.34) & 2.37 (1.01) & 2.41 (1.10) & 1.63 (2.40) & 1.63 (2.40) \\
    F & UEQ     & 27 & 1.49 (1.56) & 1.13 (1.84) & 1.59 (1.69) & 1.30 (1.35) & 1.38 (1.66) & 1.44 (1.09) \\
    F & PKI     & 27 & 1.56 (2.64) & 2.00 (1.23) & 2.63 (1.01) & 2.48 (1.11) & 1.00 (3.00) & 1.15 (2.98) \\
    M & UEQ     & 27 & 1.16 (1.33) & 0.93 (1.96) & 1.48 (1.26) & 1.25 (1.12) & 1.28 (1.11) & 1.40 (1.12) \\
    M & PKI     & 27 & 1.63 (1.78) & 2.30 (0.60) & 2.67 (0.54) & 2.48 (0.64) & 1.15 (1.90) & 1.22 (2.03) \\
    SF & UEQ    & 24 & 1.06 (1.73) & 1.06 (1.73) & 1.04 (1.30) & 1.06 (1.33) & 1.08 (1.44) & 1.48 (0.86) \\
    SF & PKI    & 24 & 1.92 (1.38) & 2.46 (0.87) & 2.75 (0.72) & 2.79 (0.26) & 0.96 (3.26) & 1.42 (2.77) \\
    SM & UEQ    & 25 & 0.55 (1.36) & 0.28 (1.54) & 0.66 (1.67) & 0.49 (1.01) & 0.78 (1.14) & 0.81 (1.01) \\
    SM & PKI    & 25 & 1.76 (1.61) & 2.32 (0.64) & 2.76 (0.44) & 2.60 (0.67) & 1.08 (2.99) & 1.00 (2.42) \\
    \bottomrule
\end{tabularx}

\caption{\textbf{UEQ Scale Scores.} This table presents the UEQ and UEQ PKI 
mean scores (and variance) for each scale, categorized by group.}
\label{tab:ueq:scores}
\end{table*}

\noindent
\textit{
\Cref{tab:ueq} presents the attributes,
\Cref{tab:ueq:scores} presents the UEQ scores for each scale
and \Cref{fig:ueq:itemized} presents the itemized
scores for each UEQ attribute.
For clarity,
we only include the positive attributes
in the figure.}

\subsection{UEQ Key Performance Indicator Extension (KPI)}

Please rate the importance of the following items relative to
your use of an ideal voter registration system for
online voting from 1 ``Not important at all'' to 7 ``Very Important''.

Do not consider how important these items are in your daily life,
but please focus specifically on how important they are when
you interact with a voter registration system for online voting.

\begin{enumerate}
    \item The ideal voter registration system should look attractive, enjoyable, friendly and pleasant.
    \item I should perform my tasks with the ideal voter registration system fast, efficient and in a pragmatic way.
    \item The ideal voter registration system should be easy to understand, clear, simple, and easy to learn.
    \item The interaction with the ideal voter registration system should be predictable, secure and meets my expectations.
    \item Using the ideal voter registration system should be interesting, exciting and motivating.
    \item The ideal voter registration system should be innovative, inventive and creatively designed.
\end{enumerate}

\subsection{Test Credentials}
\label{apx:survey:credentials}

\begin{enumerate}
    \item Did you notice anything odd while creating your real credential? Please explain.
    \item Do you remember what a test credential is? Please list the potential uses for a test credential.
    \item Did you create a test credential?
    \item After creating and marking the credentials as instructed, how confident are you that you can remember which credential is your real one?
    \item If online voting was an option today to cast votes in political elections, do you imagine yourself creating a test credential alongside your real credential? Why or why not?
    \item If online voting was an option today to cast votes in political elections, do you image yourself watching an instructional video like the one you watched before completing voter registration? Why or why not?
    \item If a government official asks you to leave all of your personal devices in a locker before entering the booth and then retrieve them after you leave the booth, would you be comfortable with this?
\end{enumerate}

\subsection{Voting Methods}
\label{apx:survey:voting-methods}
Please rate your level of trust (e.g., reliable and accurate counting and reporting of votes)
with each of the following voting options from
1 ``Not trustworthy at all'' to 7 ``Completely trustworthy''
and explain your reasoning.

\begin{enumerate}
    \item Voting in-person at a polling place, and filling out a paper ballot by hand.
    \item Voting in-person at a polling place, and submitting your choices on an electronic kiosk which then prints a ballot for you to inspect before depositing the ballot in the ballot box.
    \item Voting in-person at a polling place, and submitting your choices on an electronic kiosk.
    \item Voting remotely by filling out a paper ballot, and sending it in the mail.
    \item Voting online using an e-voting system that you also register for online.
    \item Voting online using an e-voting system, and registering for it in-person using a process similar to the one you just experienced.
\end{enumerate}
\noindent
\textit{\Cref{fig:voting-methods} presents these results.}

\subsection{Coercion Scenario}
\label{apx:survey:coercion-scenario}

\begin{enumerate}
    \item Which of the following scenarios do you think
    might plausibly happen (or have happened)
    to you or someone you know?
    Please rate the following scenarios from 1
    ``Not likely at all'' to 7 ``Extremely likely''
    \begin{itemize}
        \item Someone - such as an authority figure, employer,
        or domestic partner - coercing voters to vote in a
        particular way, threatening harm if they do not.
        \item Someone coming to voters' homes offering them money
        if they register to vote by mail and give away their
        mail-in ballot when it arrives.
        \item Someone on the Internet offering money to voters
        to record themselves voting a particular way,
        for example by taking a ``ballot selfie''.
        \item An e-voting application that sends voters money
        if they cast votes that the app instructs.
    \end{itemize}
    \item From whom do you think vote buying or coercion is most likely to happen (or have happened) to you or someone you know? Please rate the following scenarios from 1 "Not likely at all" to 7 ``Extremely likely''
    \begin{itemize}
        \item A party operative pressures or
        coerces you to vote a particular way or buys your vote.
        \item An authority (e.g., police, municipal workers)
        pressuring or coercing voters to cast a vote
        in a particular way or buys your vote.
        \item A family member, such as a domestic partner,
        pressures or coerces you to vote a particular way
        or buys your vote.
        \item An employer pressures or coerces you
        to vote a particular way or buys your vote.
    \end{itemize}
    \item For each of the above questions (1) and (2), we asked participants 
    whether this item ``Has happened to you or someone you know?''
    \begin{itemize}
        \item Yes
        \item No
        \item Describe
    \end{itemize}
    \item Any additional comments about this user study?
\end{enumerate}

\newpage

\end{document}